\documentclass[useAMS,usenatbib]{mnras}
\usepackage{graphicx,amsmath,float,amssymb,bm,xcolor}
\usepackage{rotating,lscape,url}
\def\mean#1{\left< #1 \right>}

\title[Transitional SNe Ia 2015bp and iPTF13ebh]{SN 2015bp: adding to the growing population of transitional type Ia 
supernovae}

\author[Srivastav et al.]{Shubham Srivastav$^{1,2}$\thanks{E-mail : ssrivastav@iiap.res.in}, 
G.~C.\ Anupama$^1$\thanks{E-mail : gca@iiap.res.in}, D.~K.\ Sahu$^1$, C.~D.\ Ravikumar$^2$ \\
$^1$Indian Institute of Astrophysics, II Block Koramangala, Bangalore-560 034, India \\
$^2$Department of Physics, University of Calicut, Malappuram-673635, India }

\begin{document}

\maketitle

\begin{abstract}
Photometric and spectroscopic observations of type Ia supernova 2015bp are presented, spanning $\sim -6$ to $\sim +141$
days since $B$-band maximum. Also presented are unpublished HCT spectra of type Ia iPTF13ebh between $-11$ to +34 days
since $B$-band maximum.
SN 2015bp shows rapidly declining light curves with $\Delta m_{15}(B)=1.72 \pm 0.04$. The $I$-band light curve shows a 
clear secondary maximum and peaks before the $B$-band maximum, placing SN 2015bp in the transitional category of SNe Ia.
The spectral evolution of SN 2015bp resembles other transitional SNe Ia rather than 1991bg-like events. 
The C~{\sc ii} $\lambda 6580$ feature is detected in both SN 2015bp and iPTF13ebh, though it is present till the epoch of 
$B$-band maximum in the case of SN 2015bp. The velocity gradients of Si~{\sc ii} $\lambda 6355$ place SN 2015bp and 
iPTF13ebh in the FAINT subclass, whereas pseudo-equivalent widths of Si~{\sc ii} features place them in the Cool (CL) 
subclass of SNe Ia. The bolometric light curve of SN 2015bp indicates that $\sim 0.2$ M$_{\odot}$ of $^{56}$Ni was
synthesized in the explosion, with a total ejected mass of $\sim 0.9$ M$_{\odot}$, suggesting a sub-Chandrasekhar
mass white dwarf progenitor.

\end{abstract}


\section{Introduction}

Type Ia supernovae (SNe Ia) are characterized by the absence of hydrogen and helium features and presence of prominent
absorption features of singly ionized silicon, magnesium, calcium and iron in spectra near maximum 
light \citep{1997ARA&A..35..309F}. Most SNe Ia follow the width-luminosity relation \citep{1993ApJ...413L.105P},
making them powerful standardizable cosmological distance indicators. This property of SNe Ia was instrumental in the 
discovery of accelerated expansion of the universe and dark energy \citep{1998AJ....116.1009R,1999ApJ...517..565P}.
The progenitors of SNe Ia are widely believed to be accreting carbon-oxygen white dwarfs (WDs) in close binary 
systems \citep{1960ApJ...132..565H}. The explosion is thus caused by thermonuclear runaway in the degenerate 
WD \citep{1984ApJ...286..644N}. However, the nature of the companion and details of the explosion physics are still 
not clearly understood \citep{2000ARA&A..38..191H,2011NatCo...2E.350H,2013FrPhy...8..116H}. 
Different progenitor scenarios and explosion mechanisms have been proposed to explain the observed diversity in 
SNe Ia \citep[see][]{2014ARA&A..52..107M}.
Explosion of a Chandrasekhar mass WD via a delayed detonation \citep{1991A&A...245..114K}, also called deflagration to 
detonation transition (DDT), is a popular mechanism which could possibly account for most SNe Ia \citep{2007Sci...315..825M}. 
However, \citet{2014MNRAS.445.2535S} suggested that a sizeable fraction (25-50 $\%$) of SNe Ia are sub-Chandrasekhar mass
explosions. In the sub-Chandrasekhar mass regime, double detonation \citep[eg.][]{1994ApJ...423..371W,1995ApJ...452...62L} 
is the most well studied explosion mechanism. \newline
Although a majority ($\sim 70 \%$) of SNe Ia fall within the `normal' category, $\sim 18 \%$ are subluminous 1991bg-like 
events, the remaining being overluminous SN 1991T-like and peculiar SN 2002cx-like events \citep{2011MNRAS.412.1441L}.
Previous studies have noted the conspicuous dearth of `transitional' Ia events with 
$1.5 \lesssim \Delta m_{15}(B) \lesssim 1.7$ \citep[eg.][]{2006ApJ...647..501P,2016MNRAS.460.3529A}. The question remains 
whether this is due to a selection effect, or that transitional SNe Ia are intrinsically rare events at the junction
of a bimodal distribution \citep{2016MNRAS.460.3529A}. \citet{2009AJ....138.1584K} noticed a bimodality in the NIR 
luminosity of SNe Ia, wherein events whose NIR light curves peak before the $B$-band light curve were seen to show
a normal luminosity (regardless of $\Delta m_{15}(B)$), whereas those events with later NIR maxima were seen to be 
subluminous in all bands. Recently, \citet[][hereafter H15]{2015A&A...578A...9H} defined transitional SNe Ia as the class
of fast declining events whose NIR maxima precede the $B$-band maximum, as opposed to the subluminous SN 1991bg class of
events which show late NIR maxima. This is owing to events like SN 1986G \citep{1987PASP...99..592P} and 
SN 2003gs \citep{2009AJ....138.1584K}, which show Ti~{\sc ii} features in their early spectra, but they also show early NIR
maxima and their overall properties and luminosity is intermediate to normal-bright and 1991bg-like events.
Examples of well studied transitional events in the literature include 
SN 1986G \citep{1987PASP...99..592P}, SN 2003hv \citep{2009A&A...505..265L}, SN 2003gs \citep{2009AJ....138.1584K},
SN 2004eo \citep{2007MNRAS.377.1531P}, SN 2007on \citep{2011AJ....142..156S}, SN 2009an \citep{2013MNRAS.430..869S},
SN 2011iv \citep{2012ApJ...753L...5F}, SN 2012ht \citep{2014ApJ...782L..35Y} and iPTF13ebh (H15). With intermediate 
photometric and spectroscopic properties, transitional SNe Ia signify a link between normal-bright and subluminous 
1991bg-like events, suggesting a continuous range of properties and possibly a common explosion 
mechanism \citep[H15;][]{2016MNRAS.460.3529A}.

In this paper, we present the results of photometric and spectroscopic observations of transitional type Ia supernova 
SN 2015bp. SN 2015bp was discovered on 2015 March 16.49 {\sc ut} by the Catalina Real-Time Transient Survey (CRTS) as
SNhunt281 and independently by Stan Howerton (CBAT TOCP). SN 2015bp was found at the position RA $= 15^h \, 05^m \, 30^s.1$,
Dec. $= +01^\circ \, 38' \, 02''.4$, at a discovery magnitude of 19.2 in $V$ band. The transient was offset by $\sim 39$
arcsec from the S0 galaxy NGC 5839, which has a redshift of z = 0.004 (NED). It was subsequently classified as a 1991bg-like 
event by \citet{2015ATel.7251....1J} using a spectrum obtained on March 18.0 {\sc ut} with the DEIMOS spectrograph on 
Keck II. We also present unpublished spectra of transitional Ia iPTF13ebh, which showed a fast decline with 
$\Delta m_{15}(B) = 1.79$ and displayed strong carbon features in its NIR spectra (H15).

\section{Data Reduction}

\subsection{Optical Photometry}

Photometric monitoring of SN 2015bp began on 2015 March 25 using the Himalayan Faint Object Spectrograph Camera (HFOSC)
instrument mounted on the 2-m Himalayan Chandra Telescope at the Indian Astronomical Observatory (IAO) in Hanle, India.
The HFOSC is equipped with a 2K $\times$ 4K SITe CCD, of which the central 2K $\times$ 2K region was used for imaging
observations. The pixels are 15$\micron$ $\times$ 15$\micron$ each in size. The field of view in imaging mode is 
10 arcmin $\times$ 10 arcmin, with an image scale of 0.3 arcsec pixel$^{-1}$.
Optical photometry was obtained on 23 epochs spanning $-5.9$ to +140.8 d since $B$-band maximum, using the 
Bessell $UBVRI$ filters available with the HFOSC.
Landolt standard fields \citep{1992AJ....104..340L} were observed under photometric conditions on three nights in order
to calibrate the supernova field. Standard fields PG0918+029 and PG1323-086 were observed on the night of 
2015 April 04, PG2213-006 and PG0231+051 on the night of 2015 August 19, and PG0918+029, PG0942-029 and
PG1323-006 were observed on the night of 2016 April 12.

\begin{figure}
\centering
\resizebox{0.95\hsize}{!}{\includegraphics{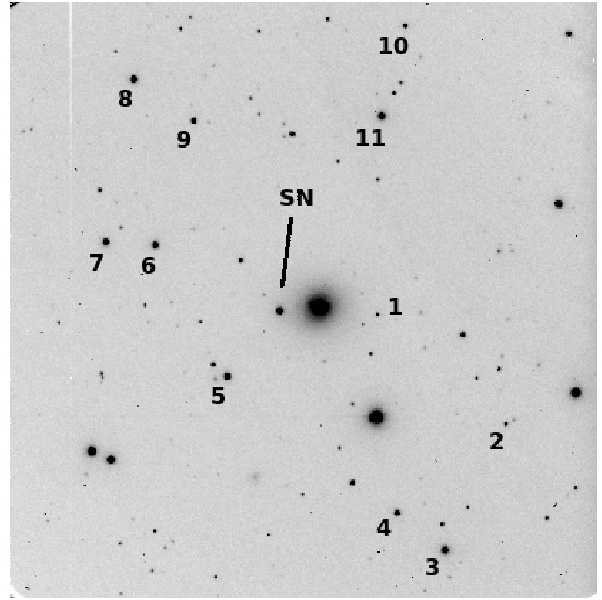}}
 \caption[]{\footnotesize Identification chart for SN 2015bp. North is up and East is to the left. The field of view is
 $10' \, \times \, 10'$. The local standards used for calibration are also indicated.}
 \label{fig:idchart}
\end{figure}

Data reduction was carried out in the standard way using various packages available in the Image Reduction and Analysis
Facility (IRAF\footnote{IRAF is distributed by the National Optical Astronomy Observatories, which are operated by the
Association of Universities for Research in Astronomy, Inc., under cooperative agreement with the National Science 
Foundation.}). 
Aperture photometry was performed on the standard stars. In order to correct for atmospheric extinction, average extinction
coefficients for the site \citep{2008BASI...36..111S} were used. Average colour terms for the HFOSC system were used
to compute the photometric zero-points during the calibration nights.
Several local standards in the supernova field were calibrated using the estimated zero-points on the calibration nights.
The local standards are identified in Figure~\ref{fig:idchart}. The $UBVRI$ magnitudes of the local standards,
averaged over three nights, are listed in Table~\ref{tab:secstd}.
Photometry was performed on the supernova and local standards using the profile-fitting method and the SN magnitudes
were derived differentially using the calibrated magnitudes of the local standards.
Nightly photometric zero-points were computed using the local standards and applied
to the supernova instrumental magnitudes. Colour corrections calculated using the local standards were incorporated
in the zero-point calculations. The summary of photometric observations and magnitudes of SN 2015bp are given in
Table~\ref{tab:mag}.

\begin{table*}
\centering
\caption{$UBVRI$ Magnitudes of local standards in the field of SN 2015bp.}
\label{tab:secstd}
\vspace{3mm} 
\begin{tabular}{c c c c c c}
 \hline 
ID & U & B & V & R & I \\ 
\hline
 1  & 16.963 $\pm$ 0.012 & 17.084 $\pm$ 0.035 & 16.860 $\pm$ 0.006 & 16.738 $\pm$ 0.022 & 16.564 $\pm$ 0.001 \\
 2  & 17.276 $\pm$ 0.028 & 17.397 $\pm$ 0.026 & 16.733 $\pm$ 0.010 & 16.350 $\pm$ 0.024 & 15.999 $\pm$ 0.014 \\
 3  & 14.337 $\pm$ 0.036 & 14.659 $\pm$ 0.024 & 14.148 $\pm$ 0.002 & 13.803 $\pm$ 0.013 & 13.471 $\pm$ 0.006 \\
 4  & 16.112 $\pm$ 0.022 & 15.925 $\pm$ 0.016 & 15.196 $\pm$ 0.019 & 14.783 $\pm$ 0.019 & 14.413 $\pm$ 0.022 \\ 
 5  & 15.176 $\pm$ 0.009 & 15.159 $\pm$ 0.019 & 14.484 $\pm$ 0.029 & 14.096 $\pm$ 0.023 & 13.722 $\pm$ 0.007 \\ 
 6  & 15.461 $\pm$ 0.006 & 15.270 $\pm$ 0.022 & 14.537 $\pm$ 0.012 & 14.142 $\pm$ 0.004 & 13.800 $\pm$ 0.003 \\ 
 7  & 15.506 $\pm$ 0.007 & 15.356 $\pm$ 0.022 & 14.648 $\pm$ 0.035 & 14.237 $\pm$ 0.029 & 13.851 $\pm$ 0.011 \\ 
 8  & 16.212 $\pm$ 0.024 & 15.656 $\pm$ 0.003 & 14.734 $\pm$ 0.012 & 14.193 $\pm$ 0.008 & 13.752 $\pm$ 0.021 \\ 
 9  & 17.880 $\pm$ 0.031 & 16.836 $\pm$ 0.015 & 15.752 $\pm$ 0.016 & 15.150 $\pm$ 0.016 & 14.655 $\pm$ 0.020 \\ 
10  & 17.210 $\pm$ 0.043 & 17.044 $\pm$ 0.018 & 16.280 $\pm$ 0.014 & 15.860 $\pm$ 0.001 & 15.448 $\pm$ 0.006 \\
11  & 15.735 $\pm$ 0.034 & 15.227 $\pm$ 0.007 & 14.291 $\pm$ 0.012 & 13.782 $\pm$ 0.007 & 13.313 $\pm$ 0.005 \\ 
 
\hline
\end{tabular}

\end{table*}

\begin{table*}
\centering
\caption{Optical $UBVRI$ photometry of SN 2015bp.}
\label{tab:mag}
\begin{tabular}{c c c c c c c c}
\hline
Date         & JD          & Phase$^*$ & U & B & V & R & I \\
(yyyy/mm/dd) & (245 7000+) & (days) & & & & & \\
\hline 
2015/03/25 & 107.48 & $-$5.85 & 13.82 $\pm$ 0.04 & 14.32 $\pm$ 0.02 & 14.24 $\pm$ 0.02 & 14.06 $\pm$ 0.02 & 14.06 $\pm$ 0.02 \\
2015/03/27 & 109.23 & $-$4.10 & 13.69 $\pm$ 0.03 & 14.12 $\pm$ 0.02 & 14.02 $\pm$ 0.02 & 13.85 $\pm$ 0.01 & 13.90 $\pm$ 0.02 \\
2015/03/31 & 113.26 & $-$0.07 & 13.63 $\pm$ 0.05 & 13.98 $\pm$ 0.03 & 13.81 $\pm$ 0.03 & 13.75 $\pm$ 0.02 & 13.93 $\pm$ 0.03 \\
2015/04/01 & 114.35 & +1.02   &                  &                  & 13.82 $\pm$ 0.02 & 13.74 $\pm$ 0.02 & 13.94 $\pm$ 0.01 \\
2015/04/02 & 115.22 & +1.89   & 13.86 $\pm$ 0.06 & 14.03 $\pm$ 0.04 & 13.82 $\pm$ 0.02 & 13.75 $\pm$ 0.02 & 13.94 $\pm$ 0.03 \\
2015/04/05 & 118.28 & +4.95   &                  &                  & 13.88 $\pm$ 0.03 & 13.90 $\pm$ 0.04 & 14.13 $\pm$ 0.02 \\
2015/04/08 & 121.32 & +7.99   & 14.49 $\pm$ 0.02 & 14.62 $\pm$ 0.02 & 14.11 $\pm$ 0.03 & 14.16 $\pm$ 0.02 & 14.31 $\pm$ 0.02 \\
2015/04/09 & 122.31 & +8.98   & 14.72 $\pm$ 0.04 & 14.79 $\pm$ 0.01 & 14.19 $\pm$ 0.03 & 14.22 $\pm$ 0.02 & 14.33 $\pm$ 0.01 \\
2015/04/10 & 123.22 & +9.89   & 14.86 $\pm$ 0.02 & 14.92 $\pm$ 0.02 & 14.26 $\pm$ 0.02 & 14.26 $\pm$ 0.02 & 14.32 $\pm$ 0.01 \\
2015/04/11 & 124.34 & +11.01  & 15.07 $\pm$ 0.02 & 15.10 $\pm$ 0.02 & 14.35 $\pm$ 0.02 & 14.30 $\pm$ 0.02 & 14.32 $\pm$ 0.02 \\
2015/04/13 & 126.32 & +12.99  & 15.39 $\pm$ 0.02 & 15.36 $\pm$ 0.02 & 14.45 $\pm$ 0.02 & 14.30 $\pm$ 0.02 & 14.24 $\pm$ 0.02 \\
2015/04/15 & 128.21 & +14.88  & 15.73 $\pm$ 0.02 & 15.68 $\pm$ 0.01 & 14.61 $\pm$ 0.01 & 14.36 $\pm$ 0.02 & 14.22 $\pm$ 0.02 \\
2015/04/17 & 130.48 & +17.15  &                  &                  & 14.74 $\pm$ 0.04 & 14.46 $\pm$ 0.02 & 14.22 $\pm$ 0.03 \\
2015/04/18 & 131.37 & +18.04  &                  & 16.09 $\pm$ 0.02 & 14.85 $\pm$ 0.02 & 14.52 $\pm$ 0.02 & 14.24 $\pm$ 0.03 \\
2015/04/24 & 137.33 & +24.00  & 16.57 $\pm$ 0.03 & 16.57 $\pm$ 0.01 & 15.38 $\pm$ 0.02 & 15.03 $\pm$ 0.02 & 14.62 $\pm$ 0.02 \\
2015/04/28 & 141.40 & +28.07  & 16.73 $\pm$ 0.02 & 16.80 $\pm$ 0.01 & 15.65 $\pm$ 0.01 & 15.33 $\pm$ 0.02 & 14.97 $\pm$ 0.02 \\
2015/04/29 & 142.20 & +28.87  & 16.79 $\pm$ 0.03 & 16.85 $\pm$ 0.02 & 15.66 $\pm$ 0.02 & 15.38 $\pm$ 0.02 & 15.03 $\pm$ 0.02 \\
2015/05/01 & 144.29 & +30.96  & 16.87 $\pm$ 0.02 & 16.91 $\pm$ 0.02 & 15.83 $\pm$ 0.03 & 15.58 $\pm$ 0.05 & 15.15 $\pm$ 0.04 \\
2015/05/05 & 148.26 & +34.93  & 17.01 $\pm$ 0.03 & 17.03 $\pm$ 0.02 & 15.94 $\pm$ 0.03 & 15.64 $\pm$ 0.02 & 15.34 $\pm$ 0.03 \\
2015/05/17 & 160.23 & +46.90  &                  & 17.33 $\pm$ 0.02 & 16.35 $\pm$ 0.01 & 16.16 $\pm$ 0.01 & 16.01 $\pm$ 0.02 \\
2015/05/22 & 165.39 & +52.06  &                  & 17.42 $\pm$ 0.01 & 16.51 $\pm$ 0.02 & 16.35 $\pm$ 0.02 & 16.25 $\pm$ 0.01 \\
2015/07/24 & 228.13 & +114.80 &                  &                  & 18.26 $\pm$ 0.03 & 18.34 $\pm$ 0.04 & 18.27 $\pm$ 0.03 \\
2015/08/19 & 254.15 & +140.82 & 19.95 $\pm$ 0.05 & 19.04 $\pm$ 0.05 & 18.70 $\pm$ 0.08 & 18.77 $\pm$ 0.10 & 18.67 $\pm$ 0.09 \\

\hline
\multicolumn{3}{l}{$^*$\footnotesize{time since $B$-band max (JD 2457113.33)}}
 \end{tabular}  
\end{table*}  

\subsection{\emph{Swift} UVOT Photometry}

SN 2015bp was followed up by the UVOT instrument \citep{2005SSRv..120...95R} on the \emph{Swift} satellite 
\citep{2004ApJ...611.1005G} from 2015 March 19 ($-$11.9 d) to 2015 May 15 (+44.4 d), where the phase denotes time since
$B$-band maximum. The images were obtained in three broadband optical filters $v$ (5468 \AA), $b$ (4392 \AA), 
$u$ (3465 \AA) and three broadband UV filters $uvw1$ (2600 \AA), $uvm2$ (2246 \AA), $uvw2$ (1928 \AA). The UVOT data for 
SN 2015bp were downloaded from the \emph{Swift} archive.
Data reduction was performed using {\sc heasoft} (High Energy Astrophysics SOFTware) following the prescriptions 
of \citet{2008MNRAS.383..627P} and \citet{2009AJ....137.4517B}. The \emph{uvotsource} task was used to extract the 
supernova magnitudes. Updated zero-points and effective area curves for the \emph{Swift} UVOT filters provided 
by \citet{2011AIPC.1358..373B} were used for the photometry.
An aperture of 5 arcsec was used for the supernova for most images except for the last few epochs,
when a smaller aperture of 3.5 arcsec was chosen for better S/N since the supernova had become quite faint.
An aperture correction prescribed by \citet{2008MNRAS.383..627P} was applied whenever the smaller aperture was used.
Sky regions of 5 arcsec each were chosen in the supernova vicinity for estimation of the background.
The estimated \emph{Swift} UVOT Vega magnitudes are listed in Table~\ref{tab:uvmag}.
The UV magnitudes of iPTF13ebh were obtained from the Swift Optical/Ultraviolet Supernova Archive
(SOUSA; \citealp{2014Ap&SS.354...89B}).

\begin{table*}
\centering
\caption{\emph{Swift} UVOT photometry of SN 2015bp.}
\label{tab:uvmag}
\vspace{3mm}
\begin{tabular}{c c c c c c c c c}
\hline
Date         & JD & Phase$^*$ & uvw2 & uvm2 & uvw1 & u & b & v \\
             & (245 7000+) & (days) &      &      &  & & & \\
\hline
2015/03/19 & 101.43 & $-$11.90 & 19.16 $\pm$ 0.19 & 17.49 $\pm$ 0.12 & 18.01 $\pm$ 0.11 & 16.15	$\pm$ 0.06 & 16.25 $\pm$ 0.05 & 15.99 $\pm$ 0.07\\
2015/03/25 & 107.09 & $-$6.24  & 16.51 $\pm$ 0.05 & 15.16 $\pm$ 0.05 & 16.72 $\pm$ 0.08 & 13.82	$\pm$ 0.04 & 14.37 $\pm$ 0.04 & 14.38 $\pm$ 0.03\\
2015/03/27 & 109.32 & $-$4.01  & 16.18 $\pm$ 0.08 & 14.90 $\pm$ 0.06 & 16.52 $\pm$ 0.09 & 13.65	$\pm$ 0.04 & 14.11 $\pm$ 0.04 & 14.02 $\pm$ 0.05\\
2015/03/30 & 111.67 & $-$1.66  & 16.23 $\pm$ 0.08 & 15.01 $\pm$ 0.06 & 16.59 $\pm$ 0.09 & 13.63	$\pm$ 0.04 & 13.89 $\pm$ 0.04 & 13.81 $\pm$ 0.05\\
2015/03/31 & 112.95 & $-$0.38  & 16.39 $\pm$ 0.07 & 15.20 $\pm$ 0.05 & 16.77 $\pm$ 0.08 & 13.75	$\pm$ 0.04 & 13.88 $\pm$ 0.04 & 13.78 $\pm$ 0.04\\
2015/04/02 & 115.41 & +2.08    & 16.68 $\pm$ 0.08 & 15.51 $\pm$ 0.06 & 16.94 $\pm$ 0.09 & 14.05	$\pm$ 0.04 & 13.95 $\pm$ 0.04 & 13.82 $\pm$ 0.04\\
2015/04/09 & 122.00 & +8.67    & 17.68 $\pm$ 0.12 & 16.20 $\pm$ 0.08 & 17.77 $\pm$ 0.12 & 15.02	$\pm$ 0.05 & 14.54 $\pm$ 0.04 & 14.07 $\pm$ 0.05\\
2015/04/12 & 124.91 & +11.58   & 17.80 $\pm$ 0.15 & 16.66 $\pm$ 0.11 & 18.39 $\pm$ 0.20 & 15.47	$\pm$ 0.07 & 15.06 $\pm$ 0.05 & 14.27 $\pm$ 0.06\\
2015/04/21 & 134.35 & +21.02   & 18.43 $\pm$ 0.19 & 17.43 $\pm$ 0.15 & 18.30 $\pm$ 0.19 & 16.56	$\pm$ 0.11 & 16.18 $\pm$ 0.07 & 15.04 $\pm$ 0.07\\
2015/04/27 & 139.90 & +26.57   & 18.94 $\pm$ 0.24 & 17.73 $\pm$ 0.16 & 18.56 $\pm$ 0.19 & 16.75	$\pm$ 0.12 & 16.66 $\pm$ 0.08 & 15.42 $\pm$ 0.08\\
2015/05/02 & 145.11 & +31.78   & 19.02 $\pm$ 0.18 & 17.82 $\pm$ 0.13 & 18.83 $\pm$ 0.17 & 17.14	$\pm$ 0.11 & 16.81 $\pm$ 0.07 & 15.79 $\pm$ 0.07\\
2015/05/07 & 150.14 & +36.81   &                  & 18.00 $\pm$ 0.11 &                  & 17.26	$\pm$ 0.10 & 16.88 $\pm$ 0.07 &                 \\
2015/05/10 & 153.30 & +39.97   & $>$18.70         & 18.12 $\pm$ 0.25 & 18.73 $\pm$ 0.23 & 17.48	$\pm$ 0.21 & 17.14 $\pm$ 0.13 & 16.07 $\pm$ 0.13\\
2015/05/15 & 157.69 & +44.36   & 19.04 $\pm$ 0.25 & 18.38 $\pm$ 0.22 & $>$19.15         & 17.57	$\pm$ 0.18 & 16.84 $\pm$ 0.08 & 16.35 $\pm$ 0.13\\

\hline
\multicolumn{3}{l}{$^*$\footnotesize{time since $B$-band max (JD 2457113.33)}}
\end{tabular}
\end{table*}

\subsection{Optical Spectroscopy}

Optical spectra of SN 2015bp were obtained on 13 epochs between $-4.1$d to +93.9d since $B$-band maximum,
while spectra of iPTF13ebh were obtained on 11 epochs spanning $-11.1$d and +33.7d since $B$-band maximum.
The observations were made using Grisms Gr7 (3500-7800 \AA) and Gr8 (5200-9250 \AA) at a spectral resolution 
of $\sim 7$ \AA. Arc lamp spectra of FeAr and FeNe were used for wavelength calibration. Spectrophotometric standard
stars HZ 44, Feige 34 and Feige 110 \citep{1990AJ.....99.1621O} were used for flux calibration. On those nights where
standard spectra were not obtained, the response curves obtained during nearby nights were used. The flux calibrated 
spectra in Gr7 and Gr8 were combined with an appropriate scale factor to obtain a single spectrum. The final spectra
were scaled to an absolute flux level using the broadband $UBVRI$ magnitudes.
For iPTF13ebh, the broadband magnitudes provided by H15 were used to scale the spectra. Telluric lines have not been 
removed from the spectra. The journal of spectroscopic observations for SN 2015bp and iPTF13ebh is shown in 
Tables~\ref{tab:speclog} and \ref{tab:speclog_13ebh}, respectively.

\begin{table*}
 \centering
 \resizebox{0.55\linewidth}{!}{%
 \begin{tabular}{c c c c}
 \hline
Date         & JD & Phase$^*$ & Range \\
(yyyy/mm/dd) & 245 7000+ & (days) & (\AA) \\
 \hline
 2015/03/27 & 109.25 & $-$4.08 & 3500-7800; 5200-9250 \\
 2015/03/31 & 113.29 & $-$0.04 & 3500-7800; 5200-9250 \\
 2015/04/02 & 115.24 & +1.91   & 3500-7800; 5200-9250 \\
 2015/04/03 & 116.22 & +2.89   & 3500-7800; 5200-9250 \\
 2015/04/05 & 118.25 & +4.92   & 3500-7800; 5200-9250 \\
 2015/04/08 & 121.33 & +8.00   & 3500-7800; 5200-9250 \\
 2015/04/09 & 122.33 & +9.00   & 3500-7800; 5200-9250 \\
 2015/04/10 & 123.24 & +9.91   & 3500-7800; 5200-9250 \\
 2015/04/15 & 128.24 & +14.91  & 3500-7800; 5200-9250 \\
 2015/04/29 & 142.23 & +28.90  & 3500-7800; 5200-9250 \\
 2015/05/03 & 146.25 & +32.92  & 3500-7800; 5200-9250 \\
 2015/05/22 & 165.40 & +52.07  & 3500-7800; 5200-9250 \\
 2015/07/03 & 207.23 & +93.90  & 3500-7800; 5200-9250 \\
 \hline
 \multicolumn{3}{l}{$^*$\footnotesize{time since $B$-band max (JD 2457113.33)}}
 \end{tabular}}
 \caption{Log of spectroscopic observations of SN 2015bp.}
 \label{tab:speclog}
\end{table*}

\begin{table*}
 \centering
 \resizebox{0.55\linewidth}{!}{%
 \begin{tabular}{c c c c}
 \hline
 Date         & JD & Phase$^*$ & Range \\
 (yyyy/mm/dd) & 245 6000+ & (days) & (\AA) \\
 \hline
 2013/11/15 & 612.29 & $-11.13$ & 3500-7800; 5200-9250 \\
 2013/11/16 & 613.37 & $-10.05$ & 3500-7800; 5200-9250 \\
 2013/11/17 & 614.31 & $-9.11$  & 3500-7800; 5200-9250 \\
 2013/11/19 & 616.29 & $-7.13$  & 3500-7800; 5200-9250 \\
 2013/11/20 & 617.34 & $-6.08$  & 3500-7800; 5200-9250 \\
 2013/11/24 & 621.35 & $-2.07$  & 3500-7800; 5200-9250 \\
 2013/11/28 & 625.37 & +1.95    & 3500-7800; 5200-9250 \\
 2013/12/04 & 631.37 & +7.95    & 3500-7800            \\
 2013/12/18 & 645.32 & +21.9    & 3500-7800; 5200-9250 \\
 2013/12/28 & 655.26 & +31.84   & 3500-7800; 5200-9250 \\
 2013/12/30 & 657.15 & +33.73   & 3500-7800; 5200-9250 \\
 \hline
 \multicolumn{3}{l}{$^*$\footnotesize{time since $B$-band max (JD 2456623.42)}}
 \end{tabular}}
 \caption{Log of spectroscopic observations of iPTF13ebh.}
 \label{tab:speclog_13ebh}
\end{table*}

\section{Photometric Analysis} \label{sec:phot}

\subsection{Light Curves}\label{subsec:lc}

\begin{figure*}
\centering
\resizebox{0.6\hsize}{!}{\includegraphics{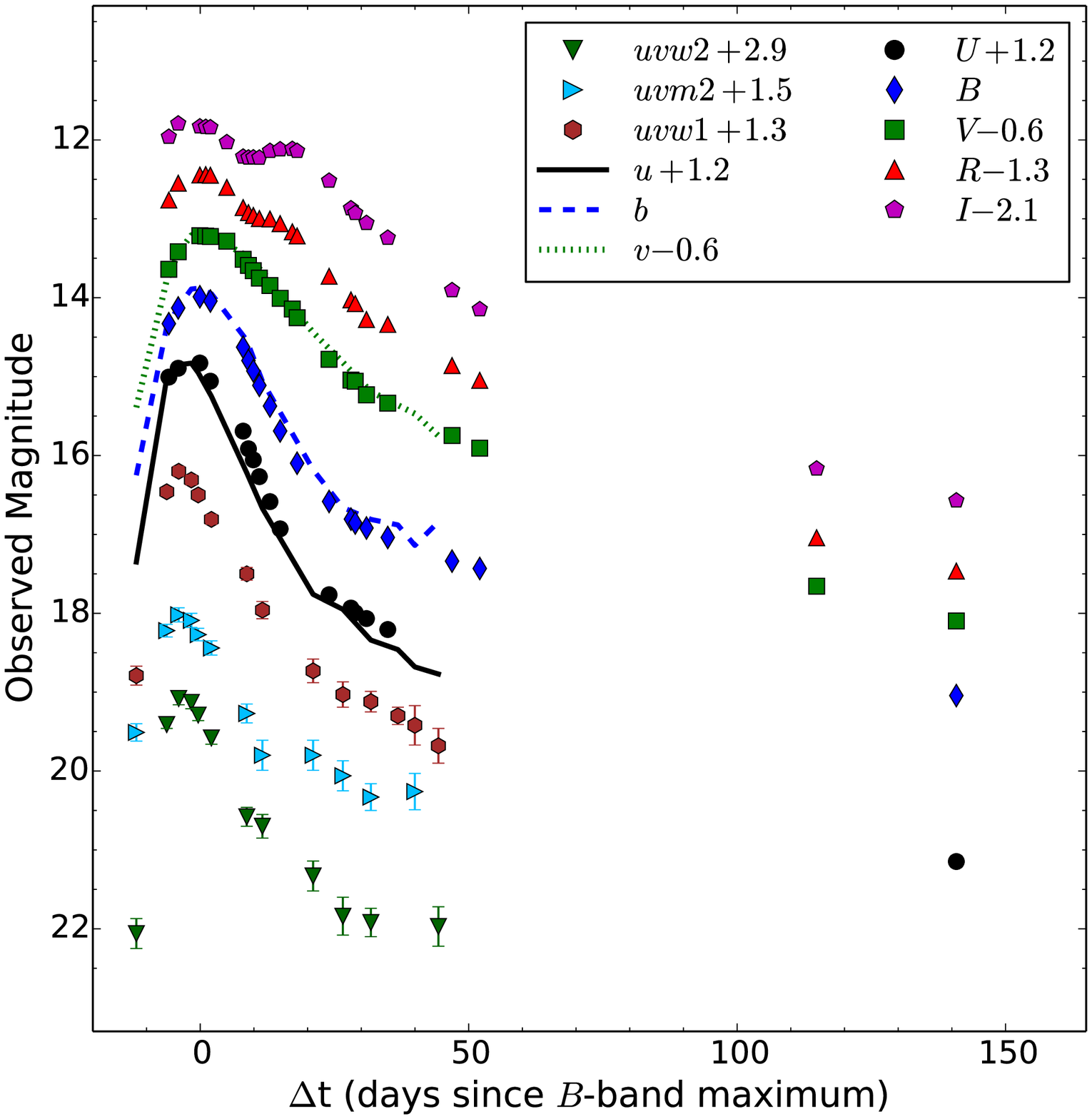}}
 \caption[]{\footnotesize HCT Optical $UBVRI$ and \emph{Swift} UV light curves of SN 2015bp. The light curves are shifted 
 along y-axis for clarity. The typical errors on $UBVRI$ magnitudes are within the symbol sizes.}
 \label{fig:lc}
\end{figure*}

\begin{figure}
\centering
\resizebox{0.95\hsize}{!}{\includegraphics{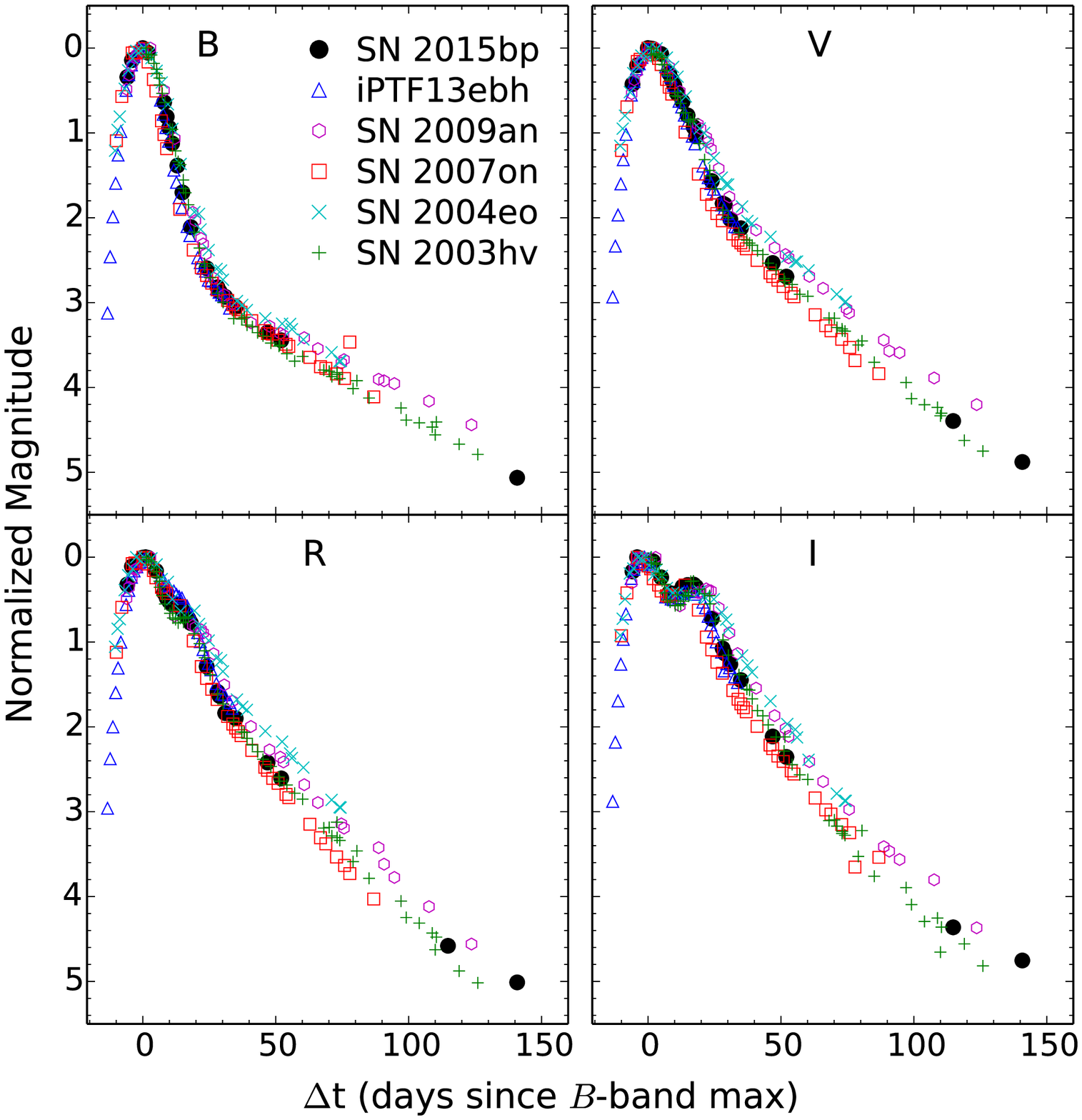}}
 \caption[]{\footnotesize Light curve comparison of SN 2015bp with iPTF13ebh (H15), SN 2009an \citep{2013MNRAS.430..869S}, 
 SN 2007on \citep{2011AJ....142..156S}, SN 2004eo \citep{2007MNRAS.377.1531P} and SN 2003hv \citep{2009A&A...505..265L}.
 The light curves of the comparison SNe have been shifted as described in the text.}
 \label{fig:lc_comp}
\end{figure}

SN 2015bp reached $B$-band maximum on 2015 March 31 (JD 2457113.33).
The HCT optical $UBVRI$ and \emph{Swift} UVOT UV light curves of SN 2015bp are shown in Figure~\ref{fig:lc}.
The light curves show a fast decline with $\Delta m_{15}(B) = 1.72 \pm 0.04$. 
However, the secondary maximum in $I$-band and shoulder in $R$-band places SN 2015bp closer to normal SNe Ia rather than
1991bg-like events.
In Figure~\ref{fig:lc_comp}, we compare the $BVRI$ light curves of SN 2015bp with a few other well studied transitional
SNe Ia which include iPTF13ebh (H15), SN 2009an \citep{2013MNRAS.430..869S}, SN 2007on \citep{2011AJ....142..156S},
SN 2004eo \citep{2007MNRAS.377.1531P} and SN 2003hv \citep{2009A&A...505..265L}.
For iPTF13ebh and SN 2007on, the published magnitudes were in $B$, $V$ and SDSS $ugri$ filters.
In order to facilitate light curve comparison, the SDSS magnitudes were converted to $R$, $I$ magnitudes using the
transformation equations provided by Lupton (2005). 
The light curves of the comparison SNe were normalized with respect to their peak magnitudes, and shifted in time
to match the epoch of their respective $B$-band maxima with that of SN 2015bp.
The $BVRI$ light curves of SN 2015bp and iPTF13ebh are quite similar, albeit iPTF13ebh shows a slightly faster
decline ($\Delta m_{15}(B) = 1.79$). 

The decline rate parameter $\Delta m_{15}(B)$ is sensitive to the amount of reddening suffered by the supernova
\citep{1999AJ....118.1766P}. For faster declining events, the $\Delta m_{15}(B)$ ceases to remain a good discriminator
between slower and faster evolving events and a reliable indicator of intrinsic colors and light curve 
shapes \citep{2014ApJ...789...32B}. This is because faster declining SNe Ia enter the linear decline phase at 
earlier times \citep{2010AJ....139..120F,2011AJ....141...19B}. \newline
The timing of the maxima in the NIR light curves relative to $B$-band is useful in distinguishing transitional
SNe Ia and 1991bg-like SNe Ia \citep{2009AJ....138.1584K}. SNe Ia which peak at later epochs in NIR bands have been found
to belong exclusively in the fast declining category ($\Delta m_{15}(B) > 1.7$) and are subluminous in all 
bands \citep{2009AJ....138.1584K,2012PASA...29..434P,2014ApJ...789...32B}.
Like in normal SNe Ia, the $I$-band maxima for SN 2015bp and iPTF13ebh (H15) occur a few days before their $B$-band 
maxima. \newline
The secondary peak in NIR light curves, occurring 20-30 days after the primary peak, is a ubiquitous feature in SNe Ia. 
The secondary peak is a consequence of a change in opacity when the ionization state in the ejecta changes due to
recombination in iron group elements \citep{2006ApJ...649..939K}.
The strength of the $I$-band secondary peak was shown to be correlated with $\Delta m_{15}(B)$ by 
\citet{1996AJ....112.2391H}, who found that faster declining events show weaker secondary maxima.
The $I$-band secondary maximum is either in the form of a weak plateau or entirely missing in subluminous SNe Ia like 
SN 1991bg \citep{1993AJ....105..301L}, SN 1998de \citep{2001PASP..113..308M}, SN 1999by \citep{2004ApJ...613.1120G}, 
SN 2005bl \citep{2008MNRAS.385...75T} and other 1991bg-like events.
In order to quantify the strength of the $I$-band secondary maximum relative to the primary, \citet{2001AJ....122.1616K}
introduced a new empirical parameter $\mean{f_{\lambda}(i)}_{20-40}$, defined as the average flux in the $I$-band between
20 to 40 days since $B$ maximum (normalized to the peak $I$-band flux). 
\citet[][Figure 6]{2014ApJ...789...32B} found a correlation between $\mean{f_{\lambda}(i)}_{20-40}$ and
$\Delta m_{15}(B)$, reaffirming the results of \citet{1996AJ....112.2391H}. However, the fast declining events 
($\Delta m_{15}(B) \gtrsim 1.7$) were seen to separate into two clusters - one with stronger
and the other with weaker secondary $I$-band maxima, respectively \citep{2014ApJ...789...32B}.
Evidently, the cluster with weaker $I$-band maxima represents the 1991bg-like subclass of events, whereas
the cluster with stronger $I$-band maxima is populated by transitional events whose overall photometric and spectroscopic
properties are not so extreme.
With $\mean{f_{\lambda}(i)}_{20-40} = 0.37$ and $0.34$, respectively, both SN 2015bp and iPTF13ebh fit in with the 
transitional cluster showing stronger $I$-band secondary maxima.

\subsection{Color Curves} \label{subsec:cc}

\begin{figure}
\centering
\resizebox{\hsize}{!}{\includegraphics{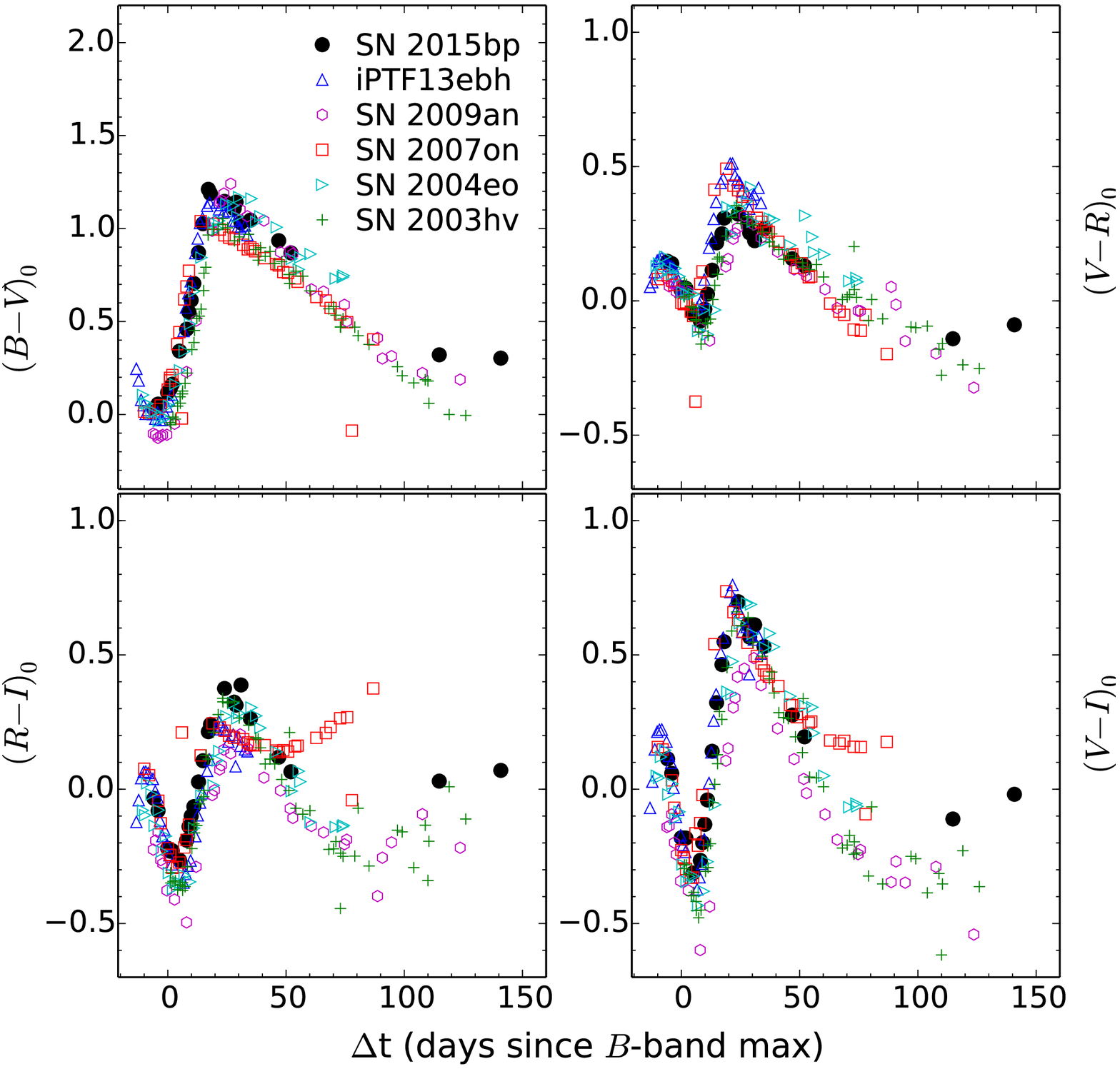}}
 \caption[]{\footnotesize Reddening corrected color curves of SN 2015bp, plotted along with color curves of transitional
 SNe Ia iPTF13ebh (H15), 2009an \citep{2013MNRAS.430..869S}, 2007on \citep{2011AJ....142..156S}, 
 2004eo \citep{2007MNRAS.377.1531P} and 2003hv \citep{2009A&A...505..265L}.}
 \label{fig:color_comp}
\end{figure}

The color curves of SNe Ia evolve towards red after maximum light due to recombination of Fe~{\sc iii} to Fe~{\sc ii} in the 
SN ejecta, which causes line blanketing in the $B$-band while increasing emissivity at longer 
wavelengths \citep{2007ApJ...656..661K}. In faster declining events with cooler ejecta, the recombination occurs at earlier
epochs, resulting in a rapid color evolution. The color evolution of SN 2015bp is shown
in Figure~\ref{fig:color_comp}, along with those of iPTF13ebh, SN 2009an, SN 2007on, SN 2004eo and SN 2003hv. 
The color curves were corrected for a Galactic reddening of $E(B-V)_{MW} = 0.046$ for SN 2015bp, $E(B-V)_{MW} = 0.068$
for iPTF13ebh, $E(B-V)_{MW} = 0.019$ for SN 2009an, $E(B-V)_{MW} = 0.01$ for SN 2007on, $E(B-V)_{MW} = 0.109$ for SN 2004eo
and $E(B-V)_{MW} = 0.016$ for SN 2003hv. In addition, a host extinction component of $E(B-V)_{host} = 0.05$ for 
iPTF13ebh (H15) and $E(B-V)_{host} = 0.089$ for SN 2009an \citep{2013MNRAS.430..869S} was also taken into account.
The reddening correction was done using a standard reddening law \citep{1989ApJ...345..245C} for $R_V = 3.1$.
The Galactic reddening values are taken from \citet{2011ApJ...737..103S}, who derived it from the dust maps of
\citet{1998ApJ...500..525S}.

At its reddest, the $(B-V)$ color curve of SN 2015bp attains a value of $1.21 \pm 0.03$, occurring $\sim 17$ days after
$B$-band maximum.
\citet{2014ApJ...789...32B} found a strong correlation between the time of $(B-V)$ maximum and $\Delta m_{15}(B)$ and
defined a new color stretch parameter, $s_{BV} = t_{max}/30$ days, where $t_{max}$ is the time of maximum for the
$(B-V)$ color curve. 
However, the correlation was seen to break down for fast decliners with
$\Delta m_{15} (B) \gtrsim 1.7$ \citep{2014ApJ...789...32B}. With $\Delta m_{15}(B) = 1.72$, SN 2015bp is at
the border of normal SNe Ia which follow the correlation between $s_{BV}$ and $\Delta m_{15}$, and the fast decliners
which do not.
The $(B-V)$ color curves of SN 2015bp ($s_{BV} = 0.57$) and iPTF13ebh ($s_{BV} = 0.63$, H15) are similar, both objects
reaching $(B-V)_{max} \sim 1.2$. SN 2007on ($s_{BV} = 0.55$) and SN 2003hv ($s_{BV} = 0.76$) attain smaller values
of $(B-V)_{max} \sim 1.05$, SN 2007on evolving slightly faster and SN 2003hv slower relative to SN 2015bp.

According to the relation provided by \citet[][Equation 3]{2014ApJ...789...32B}, the $(B-V)$ color curves of normal SNe Ia
with $\Delta m_{15} (B) \sim 1.1$ peak $\sim 30$ days after $B$ maximum. However, the $(B-V)$ color curves of
SN 2015bp and iPTF13ebh peak within 20 days of $B$ maximum. The rapid but otherwise normal color evolution of 
SN 2015bp and iPTF13ebh is consistent with their narrow, fast declining light curves.

\begin{figure}
\centering
\resizebox{0.95\hsize}{!}{\includegraphics{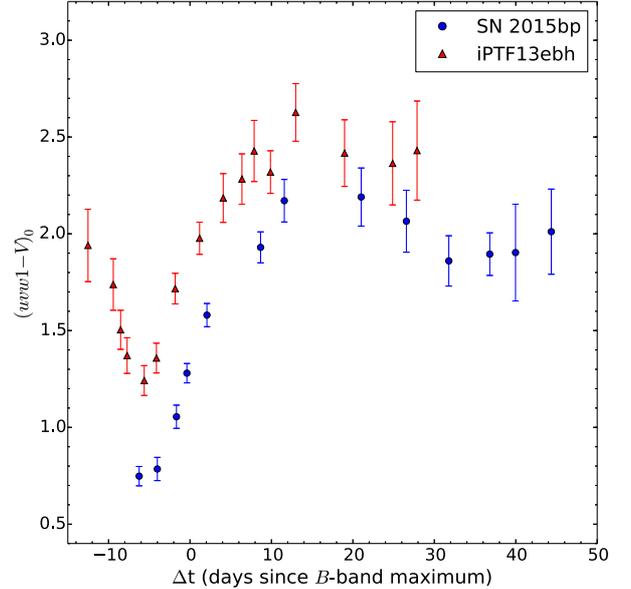}}
 \caption[]{\footnotesize Reddening corrected $(uvw1-V)$ color curves of SN 2015bp and iPTF13ebh. SN 2015bp is clearly
 bluer than iPTF13ebh, placing it in the NUV-blue group \citep{2013ApJ...779...23M}.}
 \label{fig:uvw1v}
\end{figure}

The UV-optical colors of normal SNe Ia are known to be remarkably homogeneous \citep{2010ApJ...721.1627M}.
The $(uvw1-V)$ color curve attains a blue minimum $\sim 5$ days before $B$ maximum, followed by a reddening till
$\sim 20$ days \citep{2010ApJ...721.1627M}.
Further, \citet{2013ApJ...779...23M} suggested a bimodal distribution based on the ($uvw1 - V$) color curve
into two main groups - NUV-blue and NUV-red. SN 2015bp attains a blue minimum of $(uvw1 - V)_{min} \approx 0.75$,
placing it in the NUV-blue category. Unlike SN 2015bp, iPTF13ebh is placed in the NUV-red category (H15).
In Figure~\ref{fig:uvw1v}, we show the $(uvw1-V)$ color evolution of SN 2015bp and iPTF13ebh.

\subsection{Host Reddening for SN 2015bp} \label{subsec:EBV}

There are several methods available for estimating the line of sight reddening suffered by SNe Ia in their host environment.
The slope of the $(B-V)$ color curve between 30 to 90 days since maximum light is remarkably 
homogeneous for SNe Ia which have suffered little or no reddening \citep{Lira95}. The slope of the observed $(B-V)$ color
curve can thus be used to estimate host reddening \citep{2010AJ....139..120F,2014ApJ...789...32B}.
The host reddening can also be estimated from the colors at maximum light. The difference between the observed and
expected colors around maximum light provides the color excess due to the host 
environment \citep{1999AJ....118.1766P,2004MNRAS.349.1344A,2010AJ....139..120F}.
\citet{2005ApJ...620L..87W} found a strong correlation between the $(B-V)$ color at 12 days past $B$ maximum 
($\Delta C_{12}$) and $\Delta m_{15}(B)$ for SNe Ia which have suffered low reddening due to their host.
Our light curves of SN 2015bp don't have a dense temporal coverage between 30 to 90 days since $B$-band maximum.
Thus, the colors at maximum light and $\Delta C_{12}$ have been used in order to estimate host reddening.

We derive color excesses of $E(B-V)_{host} = 0.001$, $-0.024$ and $0.058$ corresponding to the relations provided
by \citet{1999AJ....118.1766P}, \citet{2004MNRAS.349.1344A} and \citet{2010AJ....139..120F}, respectively.
The \citet{2010AJ....139..120F} relation is redder than the other two relations, with a root mean square of 0.06 mag.
From the $\Delta C_{12}$-$\Delta m_{15}(B)$ correlation of \citet{2005ApJ...620L..87W}, we derive $E(B-V)_{host} = -0.05$.
We therefore conclude that SN 2015bp has suffered very little (if any) reddening due to its host, which is consistent
with its position in the outskirts of S0 galaxy NGC 5839, and also the fact that we see no evidence of
Na~{\sc i} D absorption lines in the spectra.
We thus use $E(B-V)_{total} = E(B-V)_{MW} = 0.046$ \citep{2011ApJ...737..103S} for SN 2015bp in the subsequent analysis.

\section{Spectroscopic Analysis} \label{spec}

\subsection{Spectral Evolution and SYN++ fits} \label{subsec:spec_evol}

\begin{figure}
\centering
\resizebox{0.95\hsize}{!}{\includegraphics{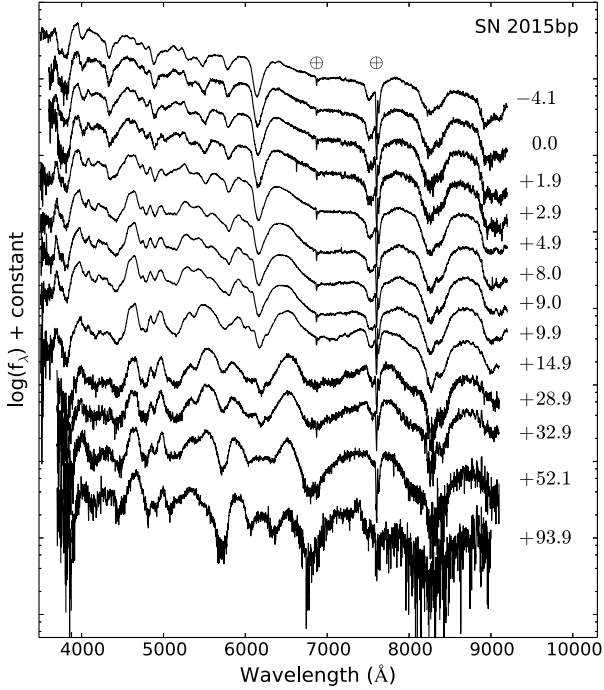}}
 \caption[]{\footnotesize Spectral time series of SN 2015bp between $-4.1$ and +93.9 days since $B$-band maximum.}
 \label{fig:spec_evol}
\end{figure}

\begin{figure}
\centering
\resizebox{0.95\hsize}{!}{\includegraphics{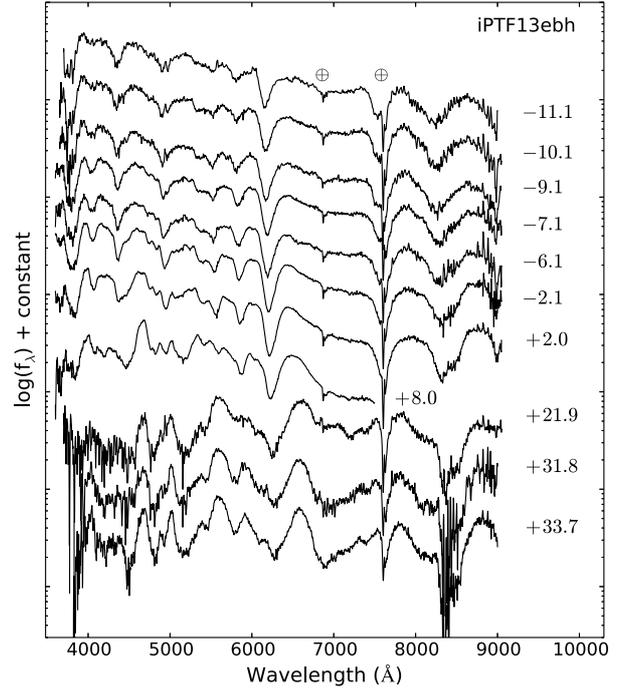}}
 \caption[]{\footnotesize Spectral time series of iPTF13ebh between $-11.1$ and +33.7 days since $B$-band maximum.}
 \label{fig:spec_evol_13ebh}
\end{figure}

The HCT spectra of SN 2015bp between $-4.1$d to +93.9d are presented here, along with unpublished HCT spectra
of iPTF13ebh between $-10.1$d to +33.7d. The $-11.1$d spectrum of iPTF13ebh was published in H15.
The spectral evolution of SN 2015bp and iPTF13ebh is shown in Figure~\ref{fig:spec_evol} and
Figure~\ref{fig:spec_evol_13ebh}, respectively.

\begin{figure}
\centering
\resizebox{0.95\hsize}{!}{\includegraphics{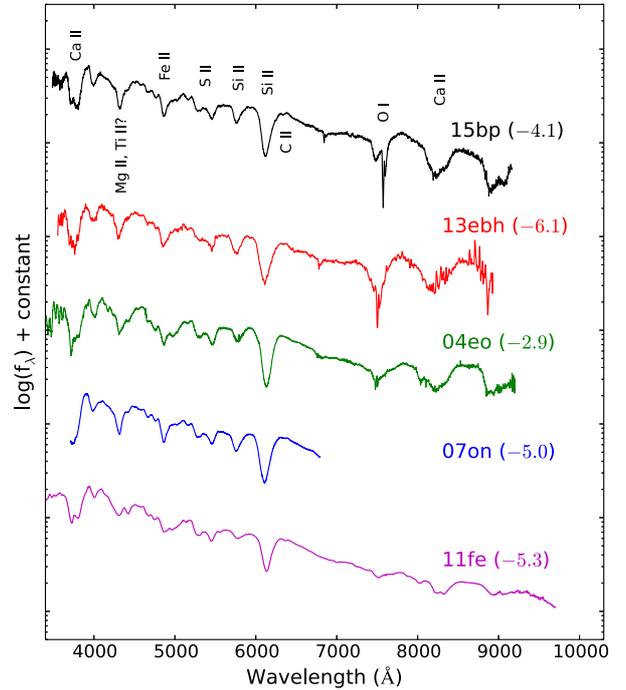}}
 \caption[]{\footnotesize Pre-maximum spectra of SN 2015bp and iPTF13ebh, compared with spectra of transitional SNe Ia
 2007on \citep{2012MNRAS.425.1789S}, 2004eo \citep{2007MNRAS.377.1531P} and normal Ia 
 SN 2011fe \citep{2013A&A...554A..27P} at similar epochs.}
 \label{fig:maxcomp1}
\end{figure}

\begin{figure}
\centering
\resizebox{0.95\hsize}{!}{\includegraphics{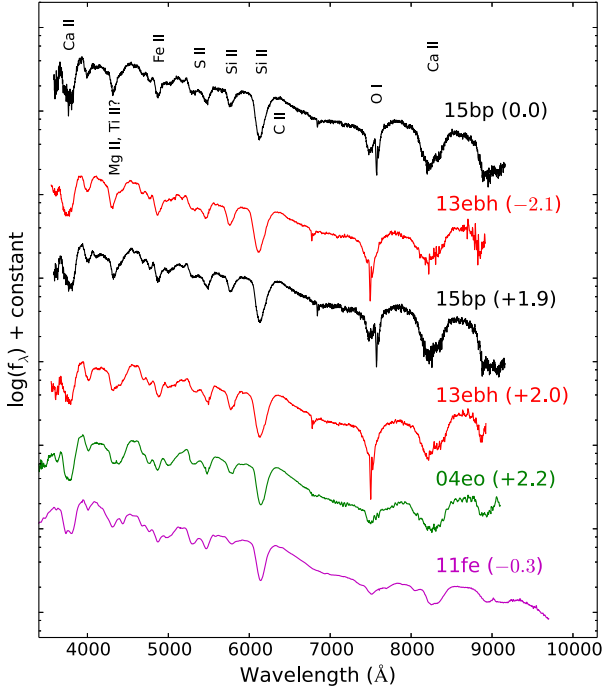}}
 \caption[]{\footnotesize Spectra of SN 2015bp and iPTF13ebh near the epoch of maximum light, compared with
 spectra of transitional SNe Ia 2007on \citep{2012MNRAS.425.1789S}, 2004eo \citep{2007MNRAS.377.1531P} and normal Ia SN 
 2011fe \citep{2013A&A...554A..27P} at similar epochs.}
 \label{fig:maxcomp2}
\end{figure}

The spectra of SN 2015bp and iPTF13ebh near the epoch of $B$-band maximum are shown in Figures~\ref{fig:maxcomp1} and 
\ref{fig:maxcomp2}, along with spectra of SN 2004eo \citep{2007MNRAS.377.1531P}, SN 2007on \citep{2012MNRAS.425.1789S} and
SN 2011fe \citep{2013A&A...554A..27P} at similar epochs for comparison.
The spectra of SN 2004eo, SN 2007on and SN 2011fe used for comparison were downloaded from the WISeREP 
archive\footnote{\url{http://wiserep.weizmann.ac.il/}} \citep{2012PASP..124..668Y}. 
The spectra were corrected for recession velocity of the respective host galaxies.
The spectra of SN 2015bp and iPTF13ebh near maximum light by and large resemble those of normal SNe Ia, with a blue
continuum and conspicuous Si~{\sc ii} absorption features. Other prominent features include S~{\sc ii}, Ca~{\sc ii},
Fe~{\sc ii}, Mg~{\sc ii} and O~{\sc i} \citep[see][]{1993ApJ...415..589K}. 
Although most of the spectral features are similar, the transitional events show a prominent absorption trough near 
4200 \AA\ attributed to Mg~{\sc ii} and Fe~{\sc iii} (H15), which is less pronounced in the normal SN 2011fe.
Also, the O~{\sc i} $\lambda 7774$ feature is stronger in the transitional and 1991bg-like events when compared to normal
SNe Ia \citep{2008MNRAS.385...75T}. \newline
Si~{\sc ii} $\lambda5972$ feature is strong relative to the Si~{\sc ii} $\lambda6355$ feature in the early
spectra for both events, which is a signature of fast declining SNe Ia \citep[eg.][]{2008MNRAS.389.1087H}.
$\mathcal{R}$(Si~{\sc ii}), which is a measure of the relative strengths of Si~{\sc ii} $\lambda5972$ and $\lambda6355$ 
features, was first introduced by \citet{1995ApJ...455L.147N} as the ratio of depths of the two features. 
Fast declining SNe Ia ($\Delta m_{15}(B) \gtrsim 1.4$) typically show higher values of 
$\mathcal{R}$(Si~{\sc ii}) ($\gtrsim 0.4$).
We measure $\mathcal{R}$(Si~{\sc ii}) of 0.55 for SN 2015bp and 0.63 for iPTF13ebh near the epoch of their respective 
$B$ maxima. The $\mathcal{R}$(Si~{\sc ii}) measurement places iPTF13ebh close to the more extreme
transitional Ia 1986G \citep{1987PASP...99..592P} and the 1991bg-like SN 2005bl \citep{2008MNRAS.385...75T}.

Pre-maximum spectra of SNe Ia often show the presence of carbon, mostly in the form of the C~{\sc ii} $\lambda 6580$ feature 
\citep[eg.][]{2011ApJ...743...27T,2011ApJ...732...30P,2012ApJ...745...74F,2012MNRAS.425.1917S}, seen as a notch redward of
the Si~{\sc ii} $\lambda 6355$ feature. This feature was first attributed to carbon by \citet{2003AJ....126.1489B}.
Detection of C~{\sc ii} $\lambda 6580$ in the early optical spectra of iPTF13ebh was reported by H15, who confirmed
the presence of carbon by detecting strong C~{\sc i} features in its NIR spectra. We detect C~{\sc ii} in
our $-11.1$d spectrum of iPTF13ebh, beyond which it becomes difficult to discern.
For SN 2015bp, we clearly detect the C~{\sc ii} $\lambda 6580$ feature in both the $-4.1$d and $0.0$d spectra.
C~{\sc ii} features usually disappear well before $B$-band maximum
\citep{2011ApJ...743...27T,2011ApJ...732...30P,2012ApJ...745...74F,2012MNRAS.425.1917S}. There are only a few examples where
C~{\sc ii} features linger till or beyond the epoch of $B$ maximum, most notably SN 2002fk \citep{2014ApJ...789...89C}, 
where C~{\sc ii} features were seen to persist $\sim 8$ days past the epoch of $B$ maximum.
The presence of optical C~{\sc ii} in SN 2015bp till maximum light and presence of strong NIR C~{\sc i} in the spectra of 
iPTF13ebh (H15) provides evidence of significant unburned material in both these events.

\begin{figure}
\centering
\resizebox{0.95\hsize}{!}{\includegraphics{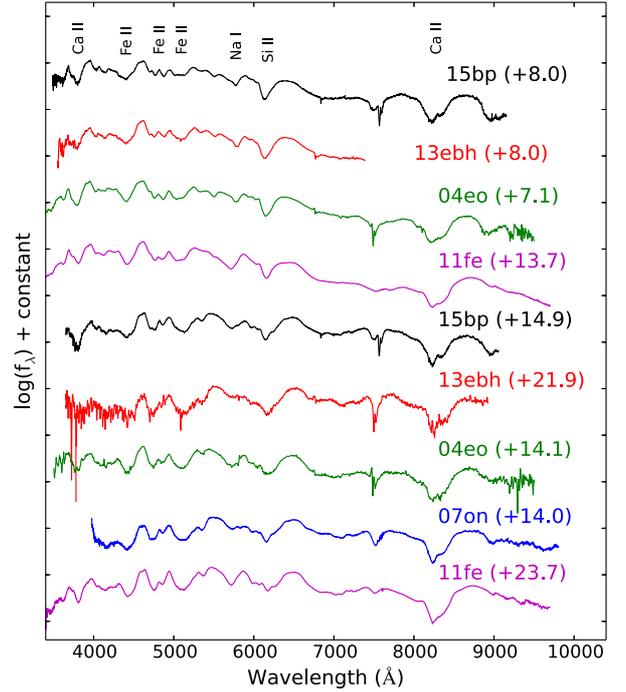}}
 \caption[]{\footnotesize Comparison of spectra of SN 2015bp and iPTF13ebh between one and three weeks since $B$ maximum
 with SN 2004eo \citep{2007MNRAS.377.1531P}, SN 2007on \citep{2012MNRAS.425.1789S} and SN 2011fe \citep{2013A&A...554A..27P}.}
 \label{fig:p15comp}
\end{figure}

\begin{figure}
\hspace*{-0.5cm}
\centering
\resizebox{0.95\hsize}{!}{\includegraphics{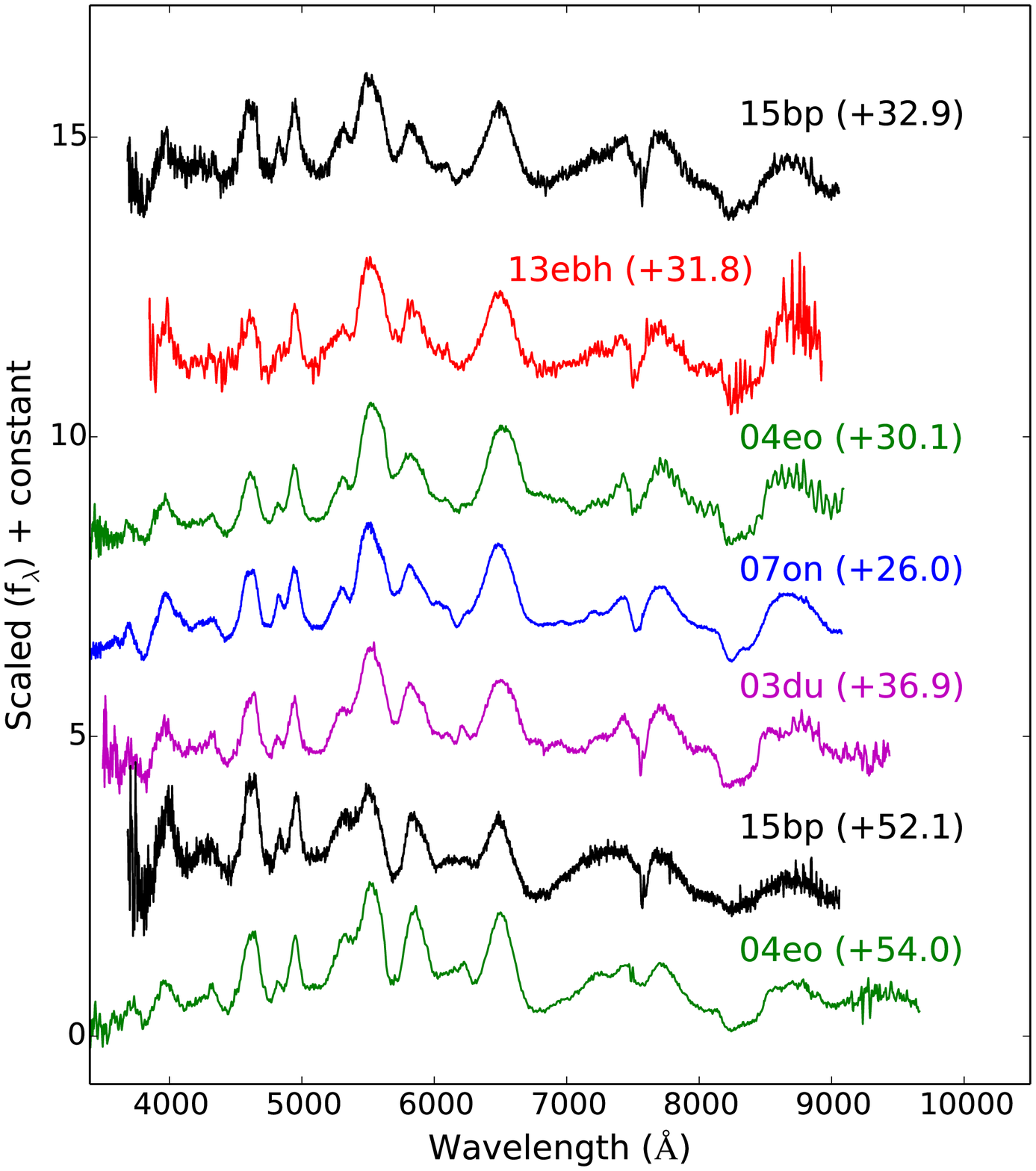}}
 \caption[]{\footnotesize Comparison of spectra of SN 2015bp and iPTF13ebh between 30 and 60 days since $B$ maximum with 
 SN 2004eo \citep{2007MNRAS.377.1531P}, SN 2007on \citep{2012MNRAS.425.1789S}, SN 2011fe \citep{2013A&A...554A..27P} and
 SN 2003du \citep{2005A&A...429..667A}.}
 \label{fig:p30comp}
\end{figure}

In Figures~\ref{fig:p15comp}-\ref{fig:p30comp}, we show the spectra of SN 2015bp and iPTF13ebh between $\sim 7-60$ days
since $B$ maximum, compared with spectra of SN 2004eo \citep{2007MNRAS.377.1531P}, SN 2007on \citep{2012MNRAS.425.1789S},
SN 2011fe \citep{2013A&A...554A..27P} and SN 2003du \citep{2005A&A...429..667A} at similar epochs. Most spectral features
are similar, with the transitional events showing a higher $\mathcal{R}$(Si~{\sc ii}) and stronger O~{\sc i} features. 
The +8.0d spectra of SN 2015bp and iPTF13ebh were seen to match well with the +13.7d spectrum of SN 2011fe. 
Similarly, the +14.9d spectrum of SN 2015bp showed a lot of similarity with the +21.7d and +23.7d spectra of SN 2011fe.
This further highlights the rapid evolution of SN 2015bp and iPTF13ebh, which is mirrored in their color evolution
(section~\ref{subsec:cc}). \newline

\begin{figure}
\centering
\resizebox{0.95\hsize}{!}{\includegraphics{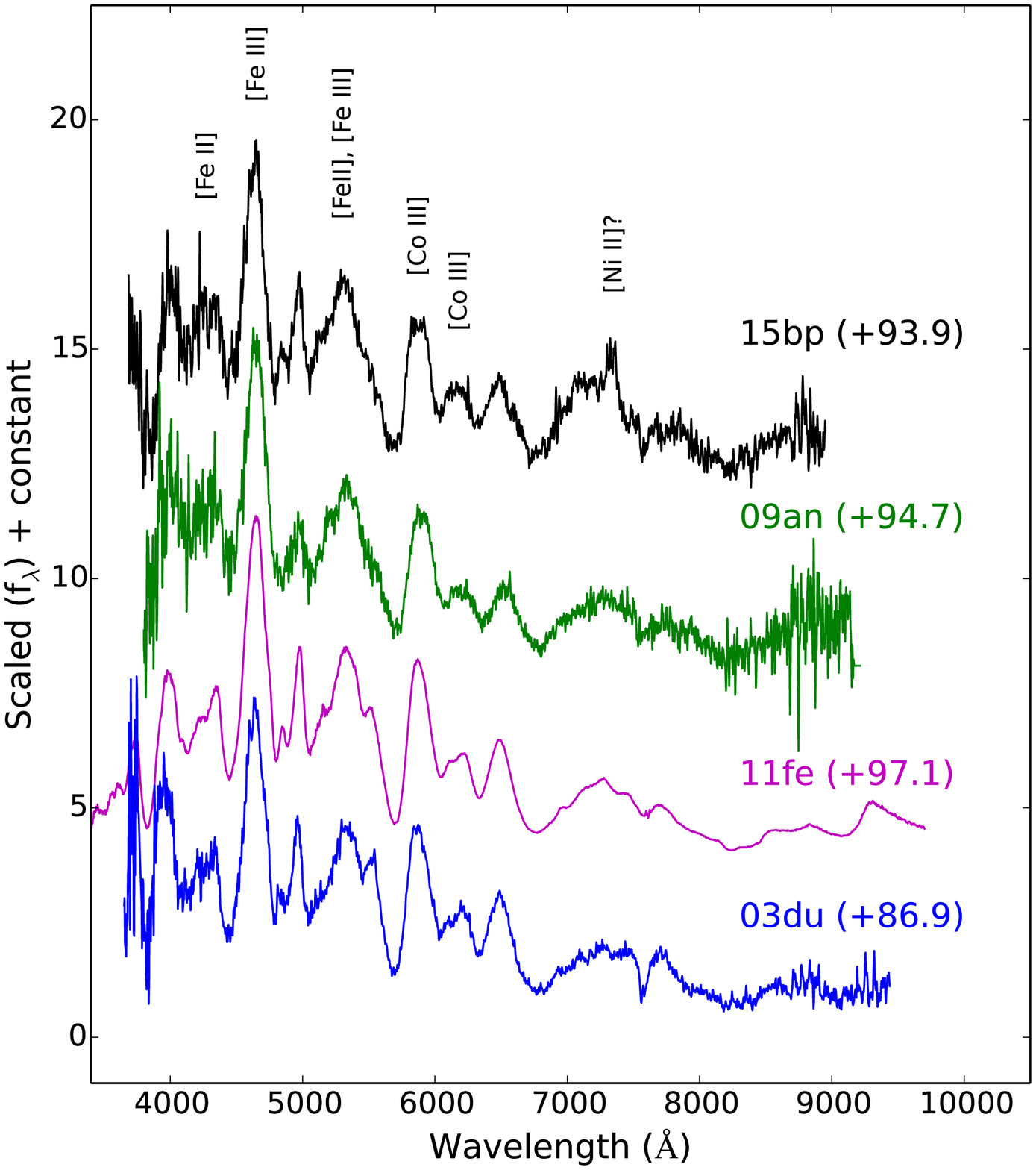}}
 \caption[]{\footnotesize Comparison of +93.9d spectrum of SN 2015bp with SN 2009an \citep{2013MNRAS.430..869S}, 
 SN 2011fe \citep{2013A&A...554A..27P} and SN 2003du \citep{2005A&A...429..667A} at similar epochs.}
 \label{fig:p90comp}
\end{figure}

Normal SNe Ia enter the nebular phase $\sim 100$ days after $B$ maximum when the ejecta becomes optically thin and can
no longer trap $\gamma$-ray photons efficiently. 
Since fast decliners show a rapid evolution, they are expected to enter the nebular phase at earlier epochs.
Figure~\ref{fig:p90comp} shows the +93.9d spectrum of SN 2015bp, plotted along with spectra of 
SN 2009an \citep{2013MNRAS.430..869S}, SN 2011fe \citep{2013A&A...554A..27P} and SN 2003du \citep{2005A&A...429..667A}
at similar epochs for comparison.
Like in the SNe Ia used for comparison, the spectrum of SN 2015bp at this phase is dominated by forbidden emission lines
[Fe~{\sc iii}] $\lambda 4701$, [Fe~{\sc ii}]/[Fe~{\sc iii}] complex $\sim 5300$ \AA\ and [Co~{\sc iii}] $\lambda 5891$,
possibly blended with [Na~{\sc i}] D lines. \newline
The width of nebular features in SNe Ia are correlated with their luminosity. Faint 1991bg-like 
events show the narrowest nebular features, whereas bright 1991T-like events show the broadest 
\citep[eg.][]{1998ApJ...499L..49M}. This can be attributed to the smaller mass of radioactive $^{56}$Ni synthesized in
1991bg-like events. For the +93.9d spectrum of SN 2015bp, we measure 
FWHM($\lambda4701$) $\sim 9300 \pm 100$ km s$^{-1}$. At a similar epoch, FWHM($\lambda 4701$) for SN 2009an is comparable
($\sim 9600 \pm 200$ km s$^{-1}$) to SN 2015bp, whereas that of the normal SNe 2011fe ($\sim 10000 \pm 100$ km s$^{-1}$) and 
2003du ($\sim 10300 \pm 100$ km s$^{-1}$) is higher, as expected.

A high blueshift ($\gtrsim 1500$ km s$^{-1}$) in the [Fe~{\sc iii}] $\lambda 4701$ feature of SNe Ia is usually seen only
in young nebular spectra obtained $< +200$ days, beyond which the central wavelength of the feature clusters around the rest
wavelength \citep{2010ApJ...708.1703M}. 
The [Fe~{\sc iii}] $\lambda 4701$ blend is blueshifted by $\sim 4000$ km s$^{-1}$ at +93.9d for SN 2015bp. 
However, velocity measurement of [Fe~{\sc iii}] $\lambda 4701$ at this early nebular epoch may be affected by P-Cygni
emission from Mg~{\sc ii} and Ti~{\sc ii}, since the transition from the optically thick to optically thin regime may not 
yet be complete.
The [Co~{\sc iii}] $\lambda 5891$ feature is also seen to be blueshifted but at a lower velocity of $\sim 1000$ km 
s$^{-1}$ at +93.9d. \citet{2016MNRAS.462..649B} reported a continuous redward shift in the central wavelength of nebular 
features, in particular [Fe~{\sc iii}] $\lambda 4701$, starting from $\sim 4600$ \AA\ at $\sim +50$ towards the rest 
wavelength till $\sim +300$ days in normal SNe Ia. This progressive shift, accompanied by a steady increase in line width,
was attributed (in part) to emergence of weak emission on the red side of [Fe~{\sc iii}] $\lambda 4701$. Unfortunately,
this trend cannot be examined further for SN 2015bp since we do not have spectra beyond +93.9d. 

Additionally, the +93.9d spectrum of SN 2015bp also seems to show a narrow emission line $\sim 7300$ \AA.
This feature could be attributed to multiple species like [Fe~{\sc ii}] $\lambda \lambda7155$, $7172$, 
[Ca~{\sc ii}] $\lambda \lambda 7291$, $7324$ or [Ni~{\sc ii}] $\lambda 7378$. However, identification of this 
feature with [Fe~{\sc ii}] or [Ca~{\sc ii}] would require it to be redshifted, whereas identification with [Ni~{\sc ii}] 
would imply a blueshift of $\sim 1400$ km s$^{-1}$. Since [Fe~{\sc iii}] and [Co~{\sc iii}] features in the spectrum show a
blueshift, we associate the feature $\sim 7300$ \AA\ with [Ni~{\sc ii}] $\lambda 7378$. 
The [Ni~{\sc ii}] $\lambda 7378$ feature is attributed to stable $^{58}$Ni which is a nuclear statistical equilibrium (NSE)
product formed in the deepest layers of SN ejecta. In the DDT scenario, stable $^{58}$Ni is created
by the initial deflagration \citep{2010ApJ...708.1703M,2011MNRAS.413.3075M}.
The [Ni~{\sc ii}] $\lambda 7378$ emission feature is usually not seen as a distinct peak till $\gtrsim 160$ days after
$B$ maximum \citep{2013MNRAS.430.1030S}. The early emergence of this feature could be explained by a rapid decrease in 
opacity in outer layers of the ejecta \citep{2009AJ....138.1584K} and a small ejecta mass which would enable escape of
$\gamma$-ray photons at earlier epochs. The small ejecta mass would also explain the fast declining light curves.
 
\begin{figure}
\centering
\resizebox{0.95\hsize}{!}{\includegraphics{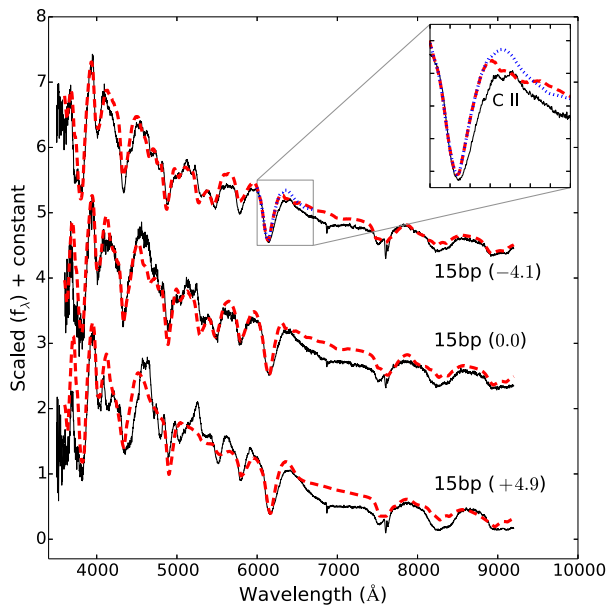}}
 \caption[]{\footnotesize Spectra of SN 2015bp near maximum light plotted along with synthetic spectra generated using
 {\sc syn++} (dashed lines). Inset shows how adding C~{\sc ii} $\lambda 6580$ (dashed line) improves the fit redward of 
 Si~{\sc ii} $\lambda 6355$, as compared to when no C~{\sc ii} is used (dotted line).}
 \label{fig:synfit}
\end{figure}

Synthetic spectra generated using {\sc syn++} \citep{2000PhDT.........6F,2011PASP..123..237T} are used to analyse the 
photospheric spectra of SN 2015bp near maximum light.
The $-4$d spectrum of SN 2015bp is fit with a photospheric velocity $v_{ph} = 11000$ km s$^{-1}$ and a blackbody 
temperature $T_{bb} = 10500$ K, whereas for the +4.9d spectrum we use $v_{ph} = 10000$ km s$^{-1}$.
The absorption trough $\sim 4200$ \AA\ is fit with strong Mg~{\sc ii}, along with a small amount 
of Ti~{\sc ii}. Inclusion of Ti~{\sc ii} improves the fit marginally in the blue wing of the $\sim 4200$ \AA\ absorption
trough \citep[eg.][]{2016MNRAS.tmp.1231A}.
High velocity (HV) components of Ca~{\sc ii} are generally seen in early spectra of most SNe Ia \citep{2005ApJ...623L..37M}.
However, like in the case of iPTF13ebh (H15), Ca~{\sc ii} HV components are not seen in SN 2015bp. 
C~{\sc ii} $\lambda 6580$ is used in the $-4.1$ and 0.0d spectra to improve the fit redward of Si~{\sc ii} $\lambda 6355$.
The C~{\sc ii} velocity was found to be comparable to the Si~{\sc ii} $\lambda 6355$ velocity.
Figure~\ref{fig:synfit} shows the {\sc syn++} fits to the early phase spectra of SN 2015bp.

\subsection{Velocity Evolution and Spectroscopic Classification} \label{subsec:vel}

Photospheric velocities of SN 2015bp and iPTF13ebh were measured using the absorption trough of 
Si~{\sc ii} $\lambda 6355$ feature. The photospheric velocity of SN 2015bp evolves from $\sim 11400$ km s$^{-1}$ at 
$-4.1$d to $\sim 9900$ km s$^{-1}$ at +14.9d; whereas for iPTF13ebh the photospheric velocity evolves from 
$\sim 13600$ km s$^{-1}$ at $-11.1$d to $\sim 8800$ km s$^{-1}$ at +21.9d.
\citet{2009ApJ...699L.139W} classified SNe Ia into two groups - Normal (N) and High-velocity (HV) on the basis of 
Si~{\sc ii} $\lambda 6355$ velocity near maximum light. HV category objects were seen to cluster $\sim 11800$ km s$^{-1}$,
whereas Normal category objects averaged $\sim 10600$ km s$^{-1}$ \citep{2009ApJ...699L.139W}.
With photospheric velocities of $\sim 11000$ km s$^{-1}$ and $\sim 10800$ km s$^{-1}$ near $B$ maximum, SN 2015bp and 
iPTF13ebh are both placed in the Normal category.

\begin{figure}
\centering
\resizebox{0.95\hsize}{!}{\includegraphics{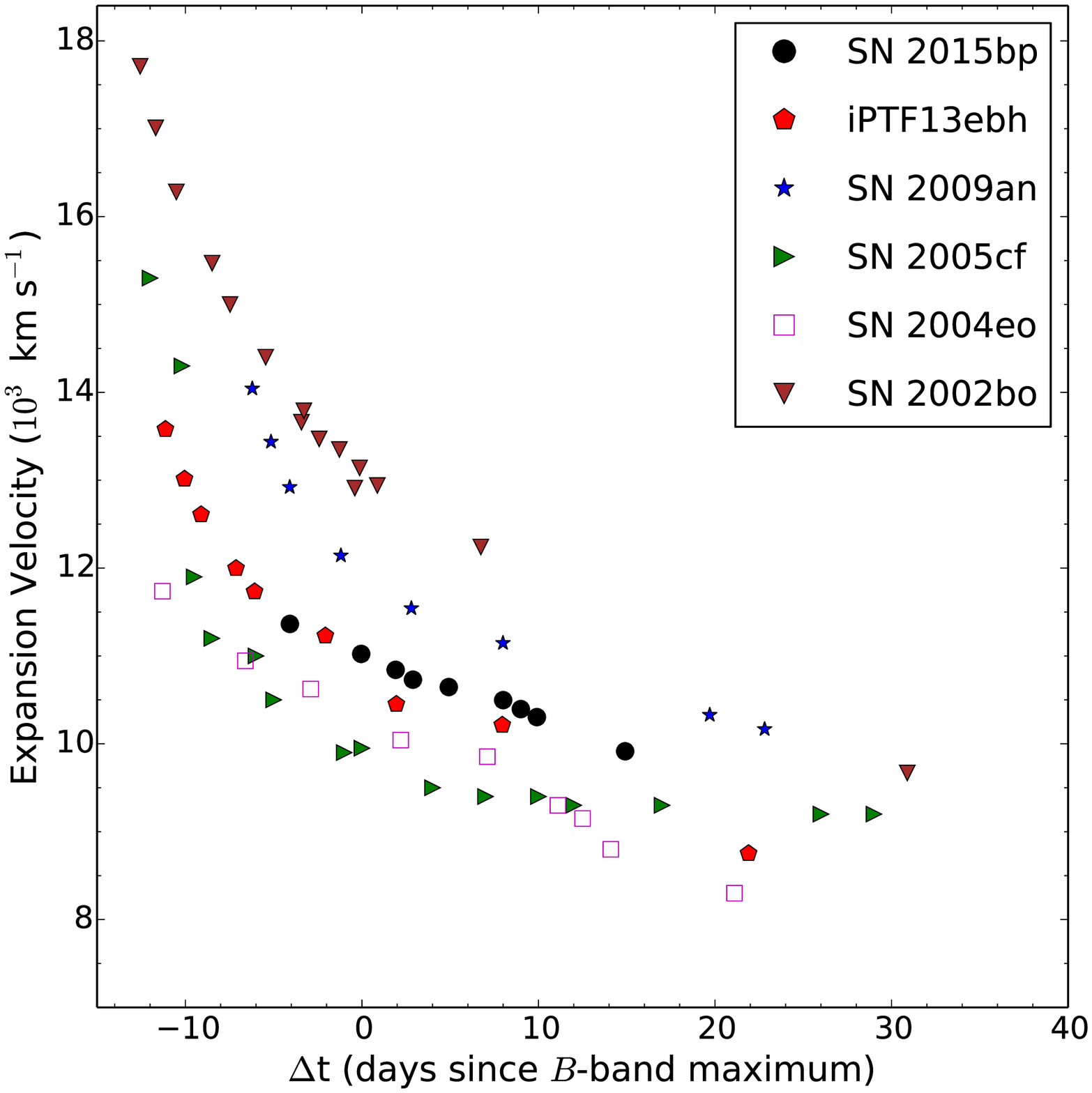}}
 \caption[]{\footnotesize Velocity evolution of Si~{\sc ii} $\lambda 6355$ for SN 2015bp and iPTF13ebh, plotted along with
 that of SN 2009an \citep{2013MNRAS.430..869S}, SN 2005cf \citep{2009ApJ...697..380W}, SN 2004eo \citep{2007MNRAS.377.1531P}
 and SN 2002bo \citep{2004MNRAS.348..261B}.}
 \label{fig:velcomp}
\end{figure}

\citet{2005ApJ...623.1011B} classified SNe Ia on the basis of the velocity gradient of Si~{\sc ii} $\lambda 6355$ feature
($\dot{v}_{Si}$). Three main subclasses of SNe Ia were found based on this definition - high velocity gradient (HVG) 
events ($\dot{v}_{Si} \gtrsim 70$ km s$^{-1}$ day$^{-1}$), low velocity gradient (LVG) events and a third subclass of
FAINT events. Objects belonging to the FAINT subclass exhibit relatively lower velocities but a rapid velocity evolution
along with fast declining light curves.
However, owing to the non-linear nature of the velocity evolution in most SNe Ia, the measured velocity gradient, and
therefore the \citet{2005ApJ...623.1011B} classification could be affected by the choice of phases between which the
velocity gradient is calculated. In addition, the Si~{\sc ii} $\lambda 6355$ feature weakens over time and suffers
increasing contamination from other neighboring features \citep{2012AJ....143..126B}.
In view of this, \citet{2012AJ....143..126B} recommended a standard definition of $\dot{v}_{Si}$ as the 
velocity gradient calculated between +0 and +10 days of $B$ maximum. \newline
The velocity evolution of Si~{\sc ii} $\lambda 6355$ feature for SN 2015bp and iPTF13ebh is plotted in 
Figure~\ref{fig:velcomp}, along with that of SN 2009an \citep{2013MNRAS.430..869S}, SN 2005cf \citep{2009ApJ...697..380W},
SN 2004eo \citep{2007MNRAS.377.1531P} and SN 2002bo \citep{2004MNRAS.348..261B}.
SN 2005cf is a LVG event whereas SN 2002bo is a HVG event. Among the transitional SNe in Figure~\ref{fig:velcomp},
SN 2009an shows consistently higher velocities than the others. The velocities of SN 2015bp and iPTF13ebh are comparable
during $B$ maximum, but iPTF13ebh shows a slightly higher velocity gradient post maximum. \newline
The velocity gradient ($\dot{v}_{Si}$) between 0 and +10 days as prescribed by
\citet{2012AJ....143..126B}, is $73 \pm 6$ km s$^{-1}$ day$^{-1}$ for SN 2015bp and $77 \pm 5$ km s$^{-1}$ for iPTF13ebh.
\citet{2005ApJ...623.1011B} reported $\left< \Delta m_{15}(B) \right> = 1.83 \pm 0.09$ and 
$\left< \mathcal{R}({\rm Si}\,\text{{\sc ii}})\right> = 0.58 \pm 0.05$ for their FAINT sample. Both SN 2015bp and 
iPTF13ebh lie in the FAINT subclass, but exhibit lower velocity gradients when compared to transitional events 
such as SN 2004eo \citep[$\sim 84$ km s$^{-1}$ day$^{-1}$]{2007MNRAS.377.1531P} and 
2009an \citep[$\sim 93$ km s$^{-1}$ day$^{-1}$]{2013MNRAS.430..869S}, and 1991bg-like events
such as SN 2005ke \citep[$\sim 125$ km s$^{-1}$ day$^{-1}$]{2012AJ....143..126B} and 
SN 2005bl \citep[$\sim 120$ km s$^{-1}$ day$^{-1}$]{2008MNRAS.385...75T}.

\begin{figure}
\centering
\resizebox{0.95\hsize}{!}{\includegraphics{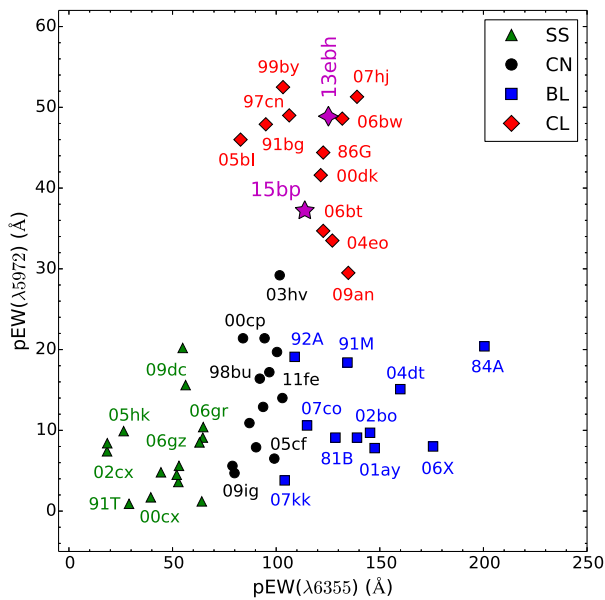}}
 \caption[]{\footnotesize Spectroscopic classification of SN 2015bp and iPTF13ebh according to the 
 \citet{2006PASP..118..560B} scheme using pEW(5972) and pEW(6355). The sample of SNe Ia is taken from 
 \citet{2006PASP..118..560B} and \citet{2012AJ....143..126B}, to which we added SN 2009an \citep{2013MNRAS.430..869S}.}
 \label{fig:pewplot}
\end{figure}

\citet{2006PASP..118..560B} constructed an alternative classification scheme wherein the ratio of 
pEWs of Si~{\sc ii} $\lambda 5972$ and Si~{\sc ii} $\lambda 6355$ was used to divide SNe Ia
into four subclasses - Core Normal (CN), Broad Line (BL), Shallow Silicon (SS) and Cool (CL).
The CN subclass is tightly clustered, forming the most homogeneous subclass \citep{2006PASP..118..560B}.
Although the pEW($\lambda5972$) of BL objects is comparable to CN objects, the BL subclass shows a broader $\lambda 6355$ 
absorption feature, i.e. higher pEW($\lambda6355$) \citep{2006PASP..118..560B}. The CL subclass usually consists of low
luminosity events with a conspicuous Si~{\sc ii} $\lambda 5972$ feature and an absorption trough near 4200 \AA\ which is 
associated with Ti~{\sc ii} features, indicating a cool ejecta \citep{2006PASP..118..560B,2012AJ....143..126B}.
By and large, the FAINT subclass of \citet{2005ApJ...623.1011B} corresponds to the CL subclass of \citet{2006PASP..118..560B};
the HVG subclass corresponds to the BL subclass and the LVG subclass includes objects belonging to both CN and SS 
subclasses \citep{2009PASP..121..238B}. \newline
The spectroscopic classification of SN 2015bp and iPTF13ebh according to the \citet{2006PASP..118..560B} scheme is shown
in Figure~\ref{fig:pewplot}. H15 measured pEW($\lambda5972$) and pEW($\lambda6355$) for iPTF13ebh as $48.9 \pm 0.6$ and
$125.2 \pm 0.5$, respectively. For SN 2015bp, the measured values of pEW($\lambda5972$) and pEW($\lambda6355$) are
$37.2 \pm 1$ and $113.8 \pm 2$. Since pEW($\lambda5972$) $>$ 30 \AA\ for both events, they are placed in the CL 
subclass \citep{2013ApJ...773...53F}.
The absorption complex $\sim 4200$ \AA\ in maximum light spectra is attributed to Mg~{\sc ii} and Fe~{\sc iii} in normal
SNe Ia. However, in 1991bg-like SNe Ia, this region is dominated by Ti~{\sc ii}. 
The pEW of this absorption region, pEW(Mg~{\sc ii}) or pEW3 $\gtrsim 220$ \AA\ for SN 1991bg-like 
events \citep{2013ApJ...773...53F}. For SN 2015bp, we measure a pEW3 of $74 \pm 2$ \AA\, whereas H15 measured a pEW3
of $104 \pm 1$ \AA\ for iPTF13ebh. Clearly, both SN 2015bp and iPTF13ebh are spectroscopically distinct from the 1991bg
class of SNe Ia. 

\section{Distance, Absolute Magnitudes and Bolometric Light Curve} \label{sec:bol}

SN 2015bp exploded in the outskirts of the S0 galaxy NGC 5839. The distance to NGC 5839 has been estimated using 
Surface Brightness Fluctuations (SBF) and Fundamental Plane (FP) methods by \citet{2013AJ....146...86T}, who reported
a distance modulus $\mu_1 = 31.78 \pm 0.17$ mag, compatible with $H_0 = 74.4$ km s$^{-1}$ Mpc$^{-1}$.
Adopting $H_0 = 72$ km s$^{-1}$ Mpc$^{-1}$ \citep{2001ApJ...553...47F}, the distance modulus becomes 
$\mu_1 = 31.85 \pm 0.17$ mag. The distance estimated using the mean Tully Fisher Relation (TFR) by \citet{2007A&A...465...71T} 
is higher, with $\mu_2 = 32.15 \pm 0.40$ mag. The corresponding distances are d$_1 = 23.44^{+1.91}_{-1.76}$ Mpc and
d$_2 = 26.92^{+5.44}_{-4.53}$ Mpc. The absolute magnitudes and bolometric properties of SN 2015bp are computed here for
the two distance scenarios d$_1$ (SBF/FP) and d$_2$ (mean TFR).

We construct the quasi-bolometric light curve of SN 2015bp using the broadband UV $uvw1, uvm2, uvw2$ and optical $UBVRI$
magnitudes presented in section~\ref{subsec:lc}. The observed magnitudes were first corrected for a total reddening of
$E(B-V) = 0.046$ (section~\ref{subsec:EBV}). 
In order to correct the UV magnitudes for reddening, we use the extinction coefficients given by \citet{2010ApJ...721.1608B}.
The $UBVRI$ extinction corrected magnitudes were converted to monochromatic fluxes using the zero-points provided by
\citet{1998A&A...333..231B}. The UV magnitudes were converted to fluxes following \citet{2008MNRAS.383..627P}.
A cubic spline function was fit through the monochromatic fluxes and integrated within appropriate wavelength limits to
obtain the UV-optical quasi-bolometric flux. 
In order to account for missing NIR flux, we use NIR corrections prescribed by \citet{2014MNRAS.440.1498S}. 
NIR corrections were calculated for SN 2015bp by averaging the predicted NIR contribution for three events from the 
sample of \citet{2014MNRAS.440.1498S} - SNF 20061020-000, SNF 20080918-004 and SN 2008ec, whose peak $B$-band
luminosity was comparable to SN 2015bp. NIR corrections were taken into account between $\sim -6$ to $\sim +52$ days.
The bolometric light curve of iPTF13ebh(1600--18000 \AA) was constructed using the UV magnitudes
from SOUSA archive \citep{2014Ap&SS.354...89B} and optical-NIR magnitudes published in H15.
For SN 2015bp, the peak UV contribution to the integrated (1600-23900 \AA) bolometric flux is $\sim 9 \%$, occurring
$\sim 4$ days before $B$ maximum. In contrast, the UV contribution to the bolometric flux for the normal SN 2013dy was 
$\sim 19 \%$ \citep{2015MNRAS.452.4307P} and $\sim 13 \%$ for SN 2011fe \citep{2013A&A...554A..27P}.
NIR contribution to bolometric flux is not expected to be high during early epochs. For SN 2004eo, NIR contribution during
maximum light was $\sim 2-3 \%$, increasing to $>20 \%$ by +50d \citep{2007MNRAS.377.1531P}. Similarly, the NIR
contribution increased from $\sim 5 \%$ at +4d to $\sim 20 \%$ at +30 for the normal SN 2005cf \citep{2009ApJ...697..380W}.
Contribution of the NIR correction applied to SN 2015bp increases from $\sim 12 \%$ near maximum to $\sim 23 \%$ at 
+30d. The extinction corrected peak apparent and absolute magnitudes of SN 2015bp for the two distance scenarios are 
tabulated in Table~\ref{tab:peakmag}.

Normal SNe Ia ($\Delta m_{15} \lesssim 1.7$) generally follow the width-luminosity relation or Phillip's 
relation \citep{1993ApJ...413L.105P}, i.e. the absolute magnitudes are correlated with the decline rate 
parameter $\Delta m_{15}$ 
\citep[eg.][]{1996AJ....112.2391H,1999AJ....118.1766P,2004MNRAS.349.1344A,2005ApJ...624..532R,2006ApJ...647..501P}.
However, the Phillip's relation shows a sudden steepening for events with $\Delta m_{15} \gtrsim 1.7$, which cannot be
accounted for by extending the linear or quadratic Phillip's relation.
\citet{2004ApJ...613.1120G} and \citet{2008MNRAS.385...75T} provided correlations between absolute magnitude and decline 
rate for fast decliners with $\Delta m_{15} \gtrsim 1.7$. Using these steeper relations, we derive $M_B^{max}$ of
$-18.30 \pm 0.20$ and $-18.40 \pm 0.20$ for SN 2015bp, respectively. \citet{2005ApJ...620L..87W} found a linear correlation
between $\Delta C_{12}$ and $M_{\lambda}^{max}$ for events with $0.81 \lesssim \Delta m_{15}(B) \lesssim 1.95$, which 
spans a wide range of peak luminosity. For a $\Delta C_{12}$ of 0.79 for SN 2015bp, the relation yields
$M_B^{max} = -18.43 \pm 0.18$. The absolute magnitudes of SN 2015bp under the d$_2$ scenario are compatible with 
these estimates (Table~\ref{tab:peakmag}). \newline
\citet{2016MNRAS.460.3529A} studied the luminosity distribution of SNe Ia in different host environments. For a sample of
16 SNe Ia in passive (E and S0) galaxies, the extinction corrected mean peak $B$ and $V$-band absolute magnitudes were 
reported as $\overline{M}_B^{max} = -18.57 \pm 0.24$ and $\overline{M}_V^{max} = -18.71 \pm 0.18$.
The peak absolute magnitudes of SN 2015bp (for both distance scenarios) are fainter than the mean values of 
\citet{2016MNRAS.460.3529A}, which is expected since the mean decline rate for their passive galaxy sample,
$\overline{\Delta m_{15}(B)} \sim 1.4$, which is lower than that of SN 2015bp. \newline
It is interesting to note here that in spite of the overall photometric and spectroscopic similarities between SN 2015bp
and iPTF13ebh, H15 reported $M_B^{max} = -18.95 \pm 0.19$ and $M_V^{max} = -19.01 \pm 0.18$ for iPTF13ebh, significantly
more luminous than SN 2015bp. iPTF13ebh falls within the Phillip's relation and becomes ``overluminous'' when compared to the
steeper relations of \citet{2004ApJ...613.1120G}, \citet{2008MNRAS.385...75T} and \citet{2005ApJ...620L..87W}.
SN 2007on \citep{2011AJ....142..156S}, which has a similar color-stretch parameter $s_{BV} = 0.55 \pm 0.02$ (H15),
also has a luminosity comparable to SN 2015bp (d$_2$ scenario) with $M_B^{max} = -18.54 \pm 0.15$ and 
$M_V^{max} = -18.67 \pm 0.15$.

\begin{table*}
 \caption{Reddening corrected peak magnitudes of SN 2015bp and predicted absolute magnitudes based on different calibrations
 of the width-luminosity relation, all of which are scaled to $H_0 = 72$ km s$^{-1}$ Mpc$^{-1}$.}
 \label{tab:peakmag}
 \vspace{3mm}
 \begin{tabular}{c c c c c c}
  \hline
 Filter & $T_{max}^*$ & Decline Rate & Peak apparent magnitude & \multicolumn{2}{|c|}{Peak absolute magnitude} \\
        & (days) & $\Delta m_{15}$     &                         & $\mu_1 = 31.85$ & $\mu_2 = 32.15$\\
  \hline
$uvw2$ & $-$3.0 & 1.67 $\pm$ 0.13 & 15.88 $\pm$ 0.07 & $-$15.97 $\pm$ 0.19 & $-$16.27 $\pm$ 0.41\\
$uvm2$ & $-$3.4 & 1.88 $\pm$ 0.15 & 16.15 $\pm$ 0.08 & $-$15.70 $\pm$ 0.19 & $-$16.00 $\pm$ 0.41\\
$uvw1$ & $-$4.1 & 1.67 $\pm$ 0.10 & 14.65 $\pm$ 0.05 & $-$17.20 $\pm$ 0.18 & $-$17.50 $\pm$ 0.40\\ 
   $U$ & $-$1.3 & 1.95 $\pm$ 0.08 & 13.35 $\pm$ 0.05 & $-$18.50 $\pm$ 0.17 & $-$18.80 $\pm$ 0.40\\ 
   $B$ &  0.0   & 1.72 $\pm$ 0.04 & 13.79 $\pm$ 0.03 & $-$18.06 $\pm$ 0.17 & $-$18.36 $\pm$ 0.40\\
   $V$ & $-$0.2 & 0.79 $\pm$ 0.04 & 13.67 $\pm$ 0.03 & $-$18.18 $\pm$ 0.17 & $-$18.48 $\pm$ 0.40\\
   $R$ & +1.0   & 0.67 $\pm$ 0.03 & 13.63 $\pm$ 0.02 & $-$18.22 $\pm$ 0.17 & $-$18.52 $\pm$ 0.40\\
   $I$ & $-$4.1 & 0.43 $\pm$ 0.04 & 13.81 $\pm$ 0.03 & $-$18.04 $\pm$ 0.17 & $-$18.34 $\pm$ 0.40\\
  \hline \hline
 \end{tabular}
 \begin{tabular}{c c c c c}
 Relation                    & $M_B^{max}$       & $M_V^{max}$       & $M_R^{max}$       & $M_I^{max}$ \\
 \hline
 \citet{2006ApJ...647..501P} & $-18.93 \pm 0.19$ & $-18.87 \pm 0.16$ & $-18.90 \pm 0.17$ & $-18.66 \pm 0.18$ \\
 \citet{2004ApJ...613.1120G} & $-18.30 \pm 0.20$ & $-18.61 \pm 0.20$ &                   & $-18.38 \pm 0.20$ \\
 \citet{2008MNRAS.385...75T} & $-18.40 \pm 0.20$ & $-18.52 \pm 0.20$ &                   &                   \\
 \citet{2005ApJ...620L..87W} & $-18.43 \pm 0.18$ & $-18.57 \pm 0.16$ &                   & $-18.42 \pm 0.16$ \\
 \hline
 \multicolumn{3}{l}{$^*$\footnotesize{time since $B$-band max (JD 2457113.33)}}
 \end{tabular}
\end{table*}


The bolometric light curves of SNe Ia are powered by the decay chain 
$^{56}$Ni $\rightarrow$ $^{56}$Co $\rightarrow$ $^{56}$Fe. Bolometric light curve shape depends on total
ejected mass, ejected $^{56}$Ni mass, explosion energy and opacity \citep{1982ApJ...253..785A,2007Sci...315..825M}.
In order to estimate the physical parameters of SN 2015bp like $^{56}$Ni mass ($M_{\rm Ni}$), total ejected 
mass ($M_{\rm ej}$) and kinetic energy of the explosion ($E_{\rm k}$), we use the Arnett--Valenti model presented by
\citet{2008MNRAS.383.1485V}. The model assumes spherical symmetry, homologous expansion of ejecta, no mixing of $^{56}$Ni, 
a small pre-explosion radius and a constant opacity ($\kappa_{\rm opt}$). The ejecta is assumed to be in the photospheric
phase, which means that the Arnett--Valenti relation is valid only for $\lesssim 30$ days past explosion. The free
parameters in the model are $M_{\rm Ni}$ and $\tau_{\rm m}$, where the latter is the time-scale of the light curve, 
defined as 
\begin{equation}\label{eq:1}
\tau_{\rm m} = \left( \frac{\kappa_{\rm opt}}{\beta c} \right)^{1/2} \left( \frac{6M^3_{\rm ej}}{5E_{\rm k}} \right)^{1/4}
\end{equation}

Here, $\beta \approx 13.8$ is a constant of integration \citep{1982ApJ...253..785A} and we use a constant
$\kappa_{\rm opt} = 0.07$ cm$^2$ g$^{-1}$ \citep[eg.][]{2016ApJ...818...79T}. For a uniform density 
\citep{1982ApJ...253..785A}, the kinetic energy can be expressed as 
\begin{equation}\label{eq:2}
 E_{\rm k} \approx \frac{3}{5}\frac{M_{\rm ej}v^2_{\rm ph}}{2}
\end{equation}
The photospheric velocity $v_{\rm ph}$ can be constrained using the spectra. Once $\tau_{\rm m}$ is obtained by fitting the
bolometric light curve to the model, Equations~(\ref{eq:1}) and (\ref{eq:2}) can be used to derive $M_{\rm ej}$ and 
$E_{\rm k}$.

\begin{figure}
\centering
\resizebox{\hsize}{!}{\includegraphics{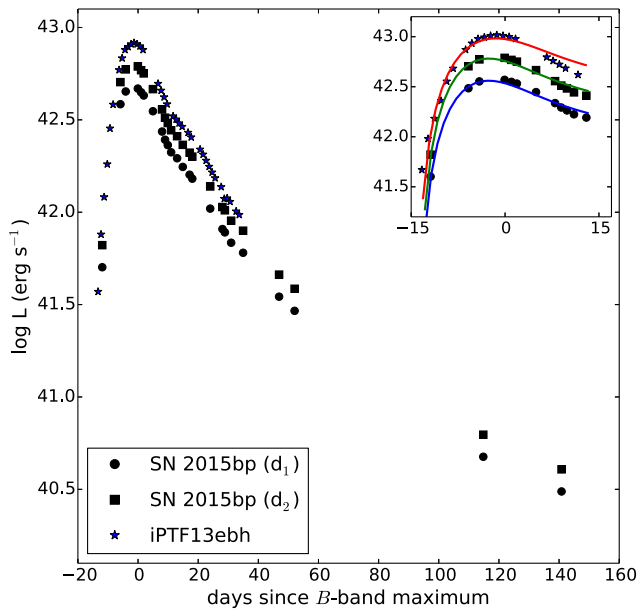}}
 \caption[]{\footnotesize Bolometric light curve of SN 2015bp(1600--23900 \AA) for two distance 
 scenarios as described in the text, plotted along with the bolometric light curve of iPTF13ebh(1600--18000 \AA). 
 Inset shows early part of the bolometric light curves of SN 2015bp and iPTF13ebh along
 with the best-fitting Arnett--Valenti model for a rise time of $\sim 14$ days for both events. 
 The bolometric light curve and model for iPTF13ebh was shifted upwards by 0.1 dex in the inset, whereas that of 
 SN 2015bp (d$_1$) was shifted downwards by 0.1 dex in the inset for clarity.}
 \label{fig:bolfit}
\end{figure}

Fitting the model to the early part of the bolometric light curve of SN 2015bp, we
obtain best-fitting parameter values of $\tau_{\rm m} = 8.79 \pm 0.54$ days, $M_{\rm Ni} = 0.15 \pm 0.01$ and $0.20 \pm 0.01$ 
M$_{\odot}$, for d$_1$ and d$_2$, respectively. Fixing $v_{\rm ph} = 11000 \pm 500$ km s$^{-1}$ as the Si~{\sc ii} velocity
near maximum, we derive $M_{\rm ej} = 0.94^{+0.17}_{-0.15}$ M$_{\odot}$ and $E_{51} = 0.68^{+0.19}_{-0.16}$ erg. 
For iPTF13ebh, the best-fitting parameters obtained are $\tau_{\rm m} = 10.26 \pm 1.05$ days and 
$M_{\rm Ni} = 0.28 \pm 0.03$ M$_{\odot}$. Fixing $v_{\rm ph} = 10800 \pm 500$ km s$^{-1}$ for iPTF13ebh, we derive 
$M_{\rm ej} = 1.26^{+0.34}_{-0.29}$ M$_{\odot}$ and $E_{51} = 0.88^{+0.34}_{-0.27}$ erg.
The bolometric light curves of SN 2015bp and iPTF13ebh are shown in Figure~\ref{fig:bolfit}, along with the best-fitting
Arnett--Valenti models.

The model favors a low $B$-band rise time of $\sim 14$ days for both SN 2015bp and iPTF13ebh.
Although normal SNe Ia have rise times of $\sim 18$ days \citep{2011MNRAS.416.2607G}, faster declining
SNe Ia are expected to have lower rise times. According to \citet{2014MNRAS.440.1498S}, the correlation between $B$-band
rise time and decline rate is given as
$$ t_{\rm R}({\rm B}) = 17.5 - 5\left(\Delta m_{15}({\rm B}) - 1.1 \right)$$
The above relation yields rise times of 14.4 and 14.1 days for SN 2015bp and iPTF13ebh, compatible with the Arnett--Valenti
model.

\section{Discussion and Conclusions}

We have presented HCT photometric and spectroscopic observations of transitional Ia SN 2015bp, along with
unpublished HCT spectra of transitional Ia iPTF13ebh. The photometric and spectroscopic properties of SN 2015bp are
intermediate to normal-bright and 1991bg-like events. \newline
Although $\Delta m_{15}(B) = 1.72$,  the $I$-band light curve shows a clear secondary maximum and peaks $\sim$ 4 days prior
to $B$ maximum. SN 2015bp is therefore placed in the transitional category along with iPTF13ebh and a few other notable 
events (see H15, table 8). \newline
As expected from the narrow light curves, SN 2015bp also shows a rapid color evolution. The $(B-V)$ color near maximum light
is $0.12 \pm 0.02$ for SN 2015bp, which is slightly redder compared to normal SNe Ia but consistent with the mean value
of $\overline{(B-V)} = 0.095 \pm 0.060$ for SNe Ia hosted by passive galaxies \citep{2016MNRAS.460.3529A}. The $(B-V)$ color
curve peaks only $\sim 17$ days after $B$ maximum ($s_{BV} = 0.57 \pm 0.03$). Although iPTF13ebh with 
$\Delta m_{15}(B) = 1.79 \pm 0.01$, $s_{BV} = 0.63 \pm 0.02$ and SN 2007on with 
$\Delta m_{15}(B) = 1.89 \pm 0.01$, $s_{BV} = 0.55 \pm 0.02$ (H15) are both faster declining than SN 2015bp, the inverse 
sequence is not seen in the color-stretch parameter $s_{BV}$. iPTF13ebh actually shows a higher value of $s_{BV}$ than 
SN 2015bp, whereas that of SN 2007on is comparable to SN 2015bp. 
iPTF13ebh, with the highest $s_{BV}$ among the three, is also the most luminous, whereas SN 2007on, with $s_{BV}$
comparable to SN 2015bp, also has a comparable luminosity within the uncertainties (section~\ref{sec:bol}).
This is further evidence that $s_{BV}$ is better suited to study fast declining events \citep{2014ApJ...789...32B}.

The $(uvw1-V)$ color of SN 2015bp places it in the NUV-blue category \citep{2010ApJ...721.1627M}. 
Although fast declining events were excluded from the analysis, \citet{2013ApJ...779...23M} found a low velocity gradient
for each NUV-blue event in their UVOT sample. Clearly, SN 2015bp does not conform to this trend. 
iPTF13ebh, which was placed in the NUV-red category by H15, is also an outlier in that it shows strong carbon features,
which were detected in only one NUV-red event in the sample of \citet{2013ApJ...779...23M}. This suggests that more
transitional events need to be studied in the UV in order to better understand the diversity within this subclass. 

Spectra of SN 2015bp also show a rapid evolution, with $\mathcal{R}$(Si~{\sc ii}) = 0.55, whereas 
$\mathcal{R}$(Si~{\sc ii}) = 0.63 for iPTF13ebh near maximum light. The presence of C~{\sc ii} $\lambda 6580$ till
the epoch of maximum light indicates a significant amount of unburned material in SN 2015bp.
The +93.9d spectrum of SN 2015bp shows an unusual emission feature $\sim 7300$ \AA, which we associate with 
Ni~{\sc ii} $\lambda 7378$ (Section~\ref{subsec:spec_evol}). The early appearance of this narrow feature points toward a 
small ejecta mass, which is consistent with our estimates of $M_{ej}$ from the bolometric light curve. 
The velocity gradient of Si~{\sc ii} $\lambda 6355$ and the pEW measurements of
Si~{\sc ii} $\lambda \lambda 6355$, $5972$ place SN 2015bp and iPTF13ebh in the CL and FAINT subclasses of
\citet{2006PASP..118..560B} and \citet{2005ApJ...623.1011B} alongside other transitional events (Section~\ref{subsec:vel}).
Although iPTF13ebh is more luminous, the $\mathcal{R}$(Si~{\sc ii}) and pEW measurements indicate that
iPTF13ebh is spectroscopically more `extreme' than SN 2015bp. H15 noted that many of the NIR spectroscopic properties of 
iPTF13ebh are similar to the 1991bg-like SN 1999by. This further highlights the diversity within the subclass of
transitional SNe Ia and emphasizes the need to study more such events in detail. 

In the d$_2$ (mean TFR) distance scenario, the absolute magnitudes $M_B^{max} = -18.36 \pm 0.40$ and 
$M_V^{max} = -18.48 \pm 0.40$ for SN 2015bp, which are consistent with the estimates from the steeper width-luminosity 
relations of \citet{2004ApJ...613.1120G} and \citet{2008MNRAS.385...75T}, and the $\Delta C_{12}$-luminosity relation
of \citet{2005ApJ...620L..87W}. Also, the d$_2$ luminosity of SN 2015bp is closer to the mean observed luminosity of SNe Ia
hosted by passive galaxies \citep{2016MNRAS.460.3529A}.
Fitting the bolometric (1600--23900 \AA) light curve of SN 2015bp with the Arnett--Valenti model indicates a $^{56}$Ni
mass $M_{\rm Ni}$ $\sim 0.2$ M$_{\odot}$, total ejected mass $M_{\rm ej}$ $\sim 0.9$ M$_{\odot}$ and kinetic
energy $E_{51}$ $\sim 0.7$ erg. The mass of $^{56}$Ni synthesized in SN 2015bp is much lower than the typical value for 
normal SNe Ia ($\sim 0.6$ M$_{\odot}$) but slightly higher than what is seen for subluminous 1991bg-like 
events ($\sim 0.1$ M$_{\odot}$).

Chandrasekhar mass DDT models are capable of producing a large range of $^{56}$Ni mass of 
$\sim 0.2-1.1$ M$_{\odot}$ \citep[eg.][]{2009Natur.460..869K,2013MNRAS.429.1156S,2013MNRAS.436..333S}, accounting for most
SNe Ia observed in nature. Double detonation models of sub-Chandrasekhar WDs can produce a similar range of $^{56}$Ni 
mass \citep[eg.][]{2010A&A...514A..53F,2010ApJ...714L..52S}. Recently, the violent merger scenario involving a double
degenerate system \citep{2010Natur.463...61P,2011A&A...528A.117P} has garnered a lot of attention. Violent mergers have been
considered as plausible progenitor scenarios for peculiar SN 2002es-like events which include 
PTF10ops \citep{2011MNRAS.418..747M}, SN 2010lp \citep{2013ApJ...778L..18K} and iPTF14atg \citep{2016MNRAS.459.4428K}.
The brightness distribution of SNe Ia produced by violent merger models is compatible with 
observations \citep{2013MNRAS.429.1425R}. Pure deflagrations of Chandrasekhar mass WDs leaving a bound 
remnant \citep[eg.][]{2012ApJ...761L..23J,2013MNRAS.429.2287K} produce weak explosions at the lower end of the 
$M_{\rm ej}-M_{\rm Ni}$ distribution for SNe Ia, whose overall properties resemble those of peculiar SN 2002cx-like
events rather than normal SNe Ia.

The $M_{\rm ej}$ and $M_{\rm Ni}$ estimates obtained for SN 2015bp in section~\ref{sec:bol} disfavour a Chandrasekhar
mass progenitor scenario. Furthermore, three-dimensional simulations of DDT models \citep{2013MNRAS.429.1156S} predict 
stable iron-group isotopes at intermediate velocities ($\sim$3000 to 10000 km s$^{-1}$), in conflict with low velocity
[Ni~{\sc ii}] detected for SN 2015bp.
For their $m_{\rm WD} = 0.97$ M$_{\odot}$ double detonation model, \citet{2010ApJ...714L..52S} 
derived $M_{\rm Ni} = 0.30$ M$_{\odot}$, $\Delta m_{15}(B) = 1.73$ and $M_{\rm B}^{\rm max} = -18.5$. The decline rate and
$B$-band luminosity observed for SN 2015bp is consistent with this model. Although the model predicts a higher $^{56}$Ni 
yield, the predicted rise time of $\sim 20$ days is also significantly higher than what was inferred for 
SN 2015bp ($\sim 14$ days). The lower inferred rise time for SN 2015bp could be attributed to presence of 
$^{56}$Ni in the outer layers of the ejecta \citep[see][]{1996ApJ...457..500H}.

An alternative for SN 2015bp could be the violent merger scenario, wherein the SN luminosity depends on the mass of the 
sub-Chandrasekhar primary CO WD undergoing prompt detonation during the merger.
For a primary WD mass of $m_{\rm WD} = 0.97$ M$_{\odot}$, \citet{2013MNRAS.429.1425R} derived a peak bolometric magnitude
of $M_{\rm bol} = -18.20$. The observed peak bolometric magnitude for SN 2015bp ($-18.0$ to $-18.3$) is consistent
with this model. Favouring a specific sub-Chandrasekhar scenario for SN 2015bp would require detailed hydrodynamic
simulations.

For iPTF13ebh, our estimates of the explosion parameters do not rule out a Chandrasekhar mass progenitor scenario.
H15 invoked a DDT model corresponding to a $^{56}$Ni mass of 0.27 M$_{\odot}$ provided by \citet{2002ApJ...568..791H}
to explain the observed properties of iPTF13ebh. Our estimates of $M_{\rm Ni} \sim 0.3$ M$_{\odot}$ and 
$M_{\rm ej} \sim 1.3$ M$_{\odot}$ are compatible with their model.

\section*{Acknowledgements}

We thank the staff of IAO, Hanle and CREST, Hosakote, that made these observations possible. The facilities at IAO
and CREST are operated by the Indian Institute of Astrophysics, Bangalore. We also thank all HCT observers
who spared part of their observing time for supernova observations. 
This work has made use of the NASA Astrophysics Data System and the NED which is operated by Jet Propulsion Laboratory, 
California Institute of Technology, under contract with the National Aeronautics and Space Administration.
This work made use of Swift/UVOT data reduced by P. J. Brown and released in the Swift Optical/Ultraviolet Supernova Archive
(SOUSA). SOUSA is supported by NASA's Astrophysics Data Analysis Program through grant NNX13AF35G. This work also made use of
the Weizmann interactive supernova data repository (WISeREP). We thank Stephane Blondin for sharing velocity gradient data
and Richard Scalzo for sharing NIR correction data.
We would also like to thank the anonymous referee whose insightful comments helped improve the quality of this manuscript.

\bibliography{biblist}

\end{document}